\documentclass[longauth,traditabstract,twocolumn]{aa}  

\usepackage{url}
\usepackage{fixltx2e}
\usepackage{ifthen}
\usepackage{epsfig}
\usepackage{graphicx}
\usepackage{epstopdf}
\usepackage{color}  
\usepackage{array}   
\usepackage{units}  
\usepackage[breaklinks,colorlinks, citecolor=blue]{hyperref}
\usepackage{natbib}  
\bibpunct{(}{)}{;}{a}{}{,} 
\usepackage{multirow}   
\usepackage{amsmath,amssymb}  
\usepackage[english]{babel}
\usepackage[latin1]{inputenc}
\usepackage[T1]{fontenc}
\usepackage{longtable,lscape}
\usepackage{footnote}

\usepackage{txfonts}
\usepackage[toc,page]{appendix} 

\providecommand{\sorthelp}[1]{}

\def\setsymbol#1#2{\expandafter\def\csname #1\endcsname{#2}}
\def\getsymbol#1{\csname #1\endcsname}

\def\Planck{\textit{Planck}}

\newbox\tablebox    \newdimen\tablewidth
\def\leaderfil{\leaders\hbox to 5pt{\hss.\hss}\hfil}
\def\endPlancktable{\tablewidth=\columnwidth 
    $$\hss\copy\tablebox\hss$$
    \vskip-\lastskip\vskip -2pt}

\def\tablenote#1 #2\par{\begingroup \parindent=0.8em
    \abovedisplayshortskip=0pt\belowdisplayshortskip=0pt
    \noindent
    $$\hss\vbox{\hsize\tablewidth \hangindent=\parindent \hangafter=1 \noindent
    \hbox to \parindent{$^#1$\hss}\strut#2\strut\par}\hss$$
    \endgroup}
\def\doubleline{\vskip 3pt\hrule \vskip 1.5pt \hrule \vskip 5pt}

\def\L2{\ifmmode L_2\else $L_2$\fi}

\def\DeltaT{\ifmmode \Delta T\else $\Delta T$\fi}
\def\deltat{\ifmmode \Delta t\else $\Delta t$\fi}
\def\fknee{\ifmmode f_{\rm knee}\else $f_{\rm knee}$\fi}
\def\Fmax{\ifmmode F_{\rm max}\else $F_{\rm max}$\fi}
\def\solar{\ifmmode{\rm M}_{\mathord\odot}\else${\rm M}_{\mathord\odot}$\fi}
\def\Msolar{\ifmmode{\rm M}_{\mathord\odot}\else${\rm M}_{\mathord\odot}$\fi}
\def\Lsolar{\ifmmode{\rm L}_{\mathord\odot}\else${\rm L}_{\mathord\odot}$\fi}
\def\inv{\ifmmode^{-1}\else$^{-1}$\fi}
\def\mo{\ifmmode^{-1}\else$^{-1}$\fi}
\def\sup#1{\ifmmode ^{\rm #1}\else $^{\rm #1}$\fi}
\def\expo#1{\ifmmode \times 10^{#1}\else $\times 10^{#1}$\fi}
\def\,{\thinspace}
\def\lsim{\mathrel{\raise .4ex\hbox{\rlap{$<$}\lower 1.2ex\hbox{$\sim$}}}}
\def\gsim{\mathrel{\raise .4ex\hbox{\rlap{$>$}\lower 1.2ex\hbox{$\sim$}}}}

\def\simprop{\mathrel{\raise .4ex\hbox{\rlap{$\propto$}\lower 1.2ex\hbox{$\sim$}}}}
\def\deg{\ifmmode^\circ\else$^\circ$\fi}
\def\pdeg{\ifmmode $\setbox0=\hbox{$^{\circ}$}\rlap{\hskip.11\wd0 .}$^{\circ}
          \else \setbox0=\hbox{$^{\circ}$}\rlap{\hskip.11\wd0 .}$^{\circ}$\fi}
\def\arcs{\ifmmode {^{\scriptstyle\prime\prime}}
          \else $^{\scriptstyle\prime\prime}$\fi}
\def\arcm{\ifmmode {^{\scriptstyle\prime}}
          \else $^{\scriptstyle\prime}$\fi}
\newdimen\sa  \newdimen\sb
\def\parcs{\sa=.07em \sb=.03em
     \ifmmode \hbox{\rlap{.}}^{\scriptstyle\prime\kern -\sb\prime}\hbox{\kern -\sa}
     \else \rlap{.}$^{\scriptstyle\prime\kern -\sb\prime}$\kern -\sa\fi}
\def\parcm{\sa=.08em \sb=.03em
     \ifmmode \hbox{\rlap{.}\kern\sa}^{\scriptstyle\prime}\hbox{\kern-\sb}
     \else \rlap{.}\kern\sa$^{\scriptstyle\prime}$\kern-\sb\fi}
\def\ra[#1 #2 #3.#4]{#1\sup{h}#2\sup{m}#3\sup{s}\llap.#4}
\def\dec[#1 #2 #3.#4]{#1\deg#2\arcm#3\arcs\llap.#4}
\def\deco[#1 #2 #3]{#1\deg#2\arcm#3\arcs}
\def\rra[#1 #2]{#1\sup{h}#2\sup{m}}

\def\dots{\relax\ifmmode \ldots\else $\ldots$\fi}
\def\WHzsr{\ifmmode $W\,Hz\mo\,sr\mo$\else W\,Hz\mo\,sr\mo\fi}
\def\mHz{\ifmmode $\,mHz$\else \,mHz\fi}
\def\GHz{\ifmmode $\,GHz$\else \,GHz\fi}
\def\mKs{\ifmmode $\,mK\,s$^{1/2}\else \,mK\,s$^{1/2}$\fi}
\def\muKs{\ifmmode \,\mu$K\,s$^{1/2}\else \,$\mu$K\,s$^{1/2}$\fi}
\def\muKRJs{\ifmmode \,\mu$K$_{\rm RJ}$\,s$^{1/2}\else \,$\mu$K$_{\rm RJ}$\,s$^{1/2}$\fi}
\def\muKHz{\ifmmode \,\mu$K\,Hz$^{-1/2}\else \,$\mu$K\,Hz$^{-1/2}$\fi}
\def\MJysr{\ifmmode \,$MJy\,sr\mo$\else \,MJy\,sr\mo\fi}
\def\MJysrmK{\ifmmode \,$MJy\,sr\mo$\,mK$_{\rm CMB}\mo\else \,MJy\,sr\mo\,mK$_{\rm CMB}\mo$\fi}
\def\microns{\ifmmode \,\mu$m$\else \,$\mu$m\fi}

\def\muK{\ifmmode \,\mu$K$\else \,$\mu$\hbox{K}\fi}
\def\microK{\ifmmode \,\mu$K$\else \,$\mu$\hbox{K}\fi}
\def\muW{\ifmmode \,\mu$W$\else \,$\mu$\hbox{W}\fi}
\def\kms{\ifmmode $\,km\,s$^{-1}\else \,km\,s$^{-1}$\fi}
\def\kmsMpc{\ifmmode $\,\kms\,Mpc\mo$\else \,\kms\,Mpc\mo\fi}
\begin{document}


\title{\textit{Planck} 2015 results. XXIII. The thermal\\
 Sunyaev-Zeldovich effect--cosmic
 infrared background correlation}

\author{\small
Planck Collaboration: P.~A.~R.~Ade\inst{82}
\and
N.~Aghanim\inst{57}
\and
M.~Arnaud\inst{71}
\and
J.~Aumont\inst{57}
\and
C.~Baccigalupi\inst{81}
\and
A.~J.~Banday\inst{89, 9}
\and
R.~B.~Barreiro\inst{62}
\and
J.~G.~Bartlett\inst{1, 64}
\and
N.~Bartolo\inst{27, 63}
\and
E.~Battaner\inst{90, 91}
% ...
}
\institute{\small
APC, AstroParticule et Cosmologie, Universit\'{e} Paris Diderot, CNRS/IN2P3, CEA/lrfu, Observatoire de Paris, Sorbonne Paris Cit\'{e}, 10, rue Alice Domon et L\'{e}onie Duquet, 75205 Paris Cedex 13, France\goodbreak
\and
African Institute for Mathematical Sciences, 6-8 Melrose Road, Muizenberg, Cape Town, South Africa\goodbreak
\and
Agenzia Spaziale Italiana Science Data Center, Via del Politecnico snc, 00133, Roma, Italy\goodbreak
\and
Aix Marseille Universit\'{e}, CNRS, LAM (Laboratoire d'Astrophysique de Marseille) UMR 7326, 13388, Marseille, France\goodbreak
\and
Astrophysics Group, Cavendish Laboratory, University of Cambridge, J J Thomson Avenue, Cambridge CB3 0HE, U.K.\goodbreak
\and
Astrophysics \& Cosmology Research Unit, School of Mathematics, Statistics \& Computer Science, University of KwaZulu-Natal, Westville Campus, Private Bag X54001, Durban 4000, South Africa\goodbreak
\and
Atacama Large Millimeter/submillimeter Array, ALMA Santiago Central Offices, Alonso de Cordova 3107, Vitacura, Casilla 763 0355, Santiago, Chile\goodbreak
\and
CITA, University of Toronto, 60 St. George St., Toronto, ON M5S 3H8, Canada\goodbreak
\and
CNRS, IRAP, 9 Av. colonel Roche, BP 44346, F-31028 Toulouse cedex 4, France\goodbreak
\and
California Institute of Technology, Pasadena, California, U.S.A.\goodbreak
\and
Centro de Estudios de F\'{i}sica del Cosmos de Arag\'{o}n (CEFCA), Plaza San Juan, 1, planta 2, E-44001, Teruel, Spain\goodbreak
\and
Computational Cosmology Center, Lawrence Berkeley National Laboratory, Berkeley, California, U.S.A.\goodbreak
\and
DSM/Irfu/SPP, CEA-Saclay, F-91191 Gif-sur-Yvette Cedex, France\goodbreak
\and
DTU Space, National Space Institute, Technical University of Denmark, Elektrovej 327, DK-2800 Kgs. Lyngby, Denmark\goodbreak
\and
D\'{e}partement de Physique Th\'{e}orique, Universit\'{e} de Gen\`{e}ve, 24, Quai E. Ansermet,1211 Gen\`{e}ve 4, Switzerland\goodbreak
\and
Departamento de Astrof\'{i}sica, Universidad de La Laguna (ULL), E-38206 La Laguna, Tenerife, Spain\goodbreak
\and
Departamento de F\'{\i}sica, Universidad de Oviedo, Avda. Calvo Sotelo s/n, Oviedo, Spain\goodbreak
\and
Department of Astrophysics/IMAPP, Radboud University Nijmegen, P.O. Box 9010, 6500 GL Nijmegen, The Netherlands\goodbreak
\and
Department of Physics \& Astronomy, University of British Columbia, 6224 Agricultural Road, Vancouver, British Columbia, Canada\goodbreak
\and
Department of Physics and Astronomy, Dana and David Dornsife College of Letter, Arts and Sciences, University of Southern California, Los Angeles, CA 90089, U.S.A.\goodbreak
\and
Department of Physics and Astronomy, University College London, London WC1E 6BT, U.K.\goodbreak
\and
Department of Physics, Florida State University, Keen Physics Building, 77 Chieftan Way, Tallahassee, Florida, U.S.A.\goodbreak
\and
Department of Physics, Gustaf H\"{a}llstr\"{o}min katu 2a, University of Helsinki, Helsinki, Finland\goodbreak
\and
Department of Physics, Princeton University, Princeton, New Jersey, U.S.A.\goodbreak
\and
Department of Physics, University of California, Santa Barbara, California, U.S.A.\goodbreak
\and
Department of Physics, University of Illinois at Urbana-Champaign, 1110 West Green Street, Urbana, Illinois, U.S.A.\goodbreak
\and
Dipartimento di Fisica e Astronomia G. Galilei, Universit\`{a} degli Studi di Padova, via Marzolo 8, 35131 Padova, Italy\goodbreak
\and
Dipartimento di Fisica e Scienze della Terra, Universit\`{a} di Ferrara, Via Saragat 1, 44122 Ferrara, Italy\goodbreak
\and
Dipartimento di Fisica, Universit\`{a} La Sapienza, P. le A. Moro 2, Roma, Italy\goodbreak
\and
Dipartimento di Fisica, Universit\`{a} degli Studi di Milano, Via Celoria, 16, Milano, Italy\goodbreak
\and
Dipartimento di Fisica, Universit\`{a} degli Studi di Trieste, via A. Valerio 2, Trieste, Italy\goodbreak
\and
Dipartimento di Matematica, Universit\`{a} di Roma Tor Vergata, Via della Ricerca Scientifica, 1, Roma, Italy\goodbreak
\and
Discovery Center, Niels Bohr Institute, Blegdamsvej 17, Copenhagen, Denmark\goodbreak
\and
European Southern Observatory, ESO Vitacura, Alonso de Cordova 3107, Vitacura, Casilla 19001, Santiago, Chile\goodbreak
\and
European Space Agency, ESAC, Planck Science Office, Camino bajo del Castillo, s/n, Urbanizaci\'{o}n Villafranca del Castillo, Villanueva de la Ca\~{n}ada, Madrid, Spain\goodbreak
\and
European Space Agency, ESTEC, Keplerlaan 1, 2201 AZ Noordwijk, The Netherlands\goodbreak
\and
Facolt\`{a} di Ingegneria, Universit\`{a} degli Studi e-Campus, Via Isimbardi 10, Novedrate (CO), 22060, Italy\goodbreak
\and
Gran Sasso Science Institute, INFN, viale F. Crispi 7, 67100 L'Aquila, Italy\goodbreak
\and
HGSFP and University of Heidelberg, Theoretical Physics Department, Philosophenweg 16, 69120, Heidelberg, Germany\goodbreak
\and
Haverford College Astronomy Department, 370 Lancaster Avenue, Haverford, Pennsylvania, U.S.A.\goodbreak
\and
Helsinki Institute of Physics, Gustaf H\"{a}llstr\"{o}min katu 2, University of Helsinki, Helsinki, Finland\goodbreak
\and
INAF - Osservatorio Astrofisico di Catania, Via S. Sofia 78, Catania, Italy\goodbreak
\and
INAF - Osservatorio Astronomico di Padova, Vicolo dell'Osservatorio 5, Padova, Italy\goodbreak
\and
INAF - Osservatorio Astronomico di Roma, via di Frascati 33, Monte Porzio Catone, Italy\goodbreak
\and
INAF - Osservatorio Astronomico di Trieste, Via G.B. Tiepolo 11, Trieste, Italy\goodbreak
\and
INAF/IASF Bologna, Via Gobetti 101, Bologna, Italy\goodbreak
\and
INAF/IASF Milano, Via E. Bassini 15, Milano, Italy\goodbreak
\and
INFN, Sezione di Bologna, Via Irnerio 46, I-40126, Bologna, Italy\goodbreak
\and
INFN, Sezione di Roma 1, Universit\`{a} di Roma Sapienza, Piazzale Aldo Moro 2, 00185, Roma, Italy\goodbreak
\and
INFN, Sezione di Roma 2, Universit\`{a} di Roma Tor Vergata, Via della Ricerca Scientifica, 1, Roma, Italy\goodbreak
\and
INFN/National Institute for Nuclear Physics, Via Valerio 2, I-34127 Trieste, Italy\goodbreak
\and
IPAG: Institut de Plan\'{e}tologie et d'Astrophysique de Grenoble, Universit\'{e} Grenoble Alpes, IPAG, F-38000 Grenoble, France, CNRS, IPAG, F-38000 Grenoble, France\goodbreak
\and
IUCAA, Post Bag 4, Ganeshkhind, Pune University Campus, Pune 411 007, India\goodbreak
\and
Imperial College London, Astrophysics group, Blackett Laboratory, Prince Consort Road, London, SW7 2AZ, U.K.\goodbreak
\and
Infrared Processing and Analysis Center, California Institute of Technology, Pasadena, CA 91125, U.S.A.\goodbreak
\and
Institut Universitaire de France, 103, bd Saint-Michel, 75005, Paris, France\goodbreak
\and
Institut d'Astrophysique Spatiale, CNRS (UMR8617) Universit\'{e} Paris-Sud 11, B\^{a}timent 121, Orsay, France\goodbreak
\and
Institut d'Astrophysique de Paris, CNRS (UMR7095), 98 bis Boulevard Arago, F-75014, Paris, France\goodbreak
\and
Institute of Astronomy, University of Cambridge, Madingley Road, Cambridge CB3 0HA, U.K.\goodbreak
\and
Institute of Theoretical Astrophysics, University of Oslo, Blindern, Oslo, Norway\goodbreak
\and
Instituto de Astrof\'{\i}sica de Canarias, C/V\'{\i}a L\'{a}ctea s/n, La Laguna, Tenerife, Spain\goodbreak
\and
Instituto de F\'{\i}sica de Cantabria (CSIC-Universidad de Cantabria), Avda. de los Castros s/n, Santander, Spain\goodbreak
\and
Istituto Nazionale di Fisica Nucleare, Sezione di Padova, via Marzolo 8, I-35131 Padova, Italy\goodbreak
\and
Jet Propulsion Laboratory, California Institute of Technology, 4800 Oak Grove Drive, Pasadena, California, U.S.A.\goodbreak
\and
Jodrell Bank Centre for Astrophysics, Alan Turing Building, School of Physics and Astronomy, The University of Manchester, Oxford Road, Manchester, M13 9PL, U.K.\goodbreak
\and
Kavli Institute for Cosmological Physics, University of Chicago, Chicago, IL 60637, USA\goodbreak
\and
Kavli Institute for Cosmology Cambridge, Madingley Road, Cambridge, CB3 0HA, U.K.\goodbreak
\and
Kazan Federal University, 18 Kremlyovskaya St., Kazan, 420008, Russia\goodbreak
\and
LAL, Universit\'{e} Paris-Sud, CNRS/IN2P3, Orsay, France\goodbreak
\and
LERMA, CNRS, Observatoire de Paris, 61 Avenue de l'Observatoire, Paris, France\goodbreak
\and
Laboratoire AIM, IRFU/Service d'Astrophysique - CEA/DSM - CNRS - Universit\'{e} Paris Diderot, B\^{a}t. 709, CEA-Saclay, F-91191 Gif-sur-Yvette Cedex, France\goodbreak
\and
Laboratoire de Physique Subatomique et Cosmologie, Universit\'{e} Grenoble-Alpes, CNRS/IN2P3, 53, rue des Martyrs, 38026 Grenoble Cedex, France\goodbreak
\and
Laboratoire de Physique Th\'{e}orique, Universit\'{e} Paris-Sud 11 \& CNRS, B\^{a}timent 210, 91405 Orsay, France\goodbreak
\and
Lawrence Berkeley National Laboratory, Berkeley, California, U.S.A.\goodbreak
\and
Lebedev Physical Institute of the Russian Academy of Sciences, Astro Space Centre, 84/32 Profsoyuznaya st., Moscow, GSP-7, 117997, Russia\goodbreak
\and
Max-Planck-Institut f\"{u}r Astrophysik, Karl-Schwarzschild-Str. 1, 85741 Garching, Germany\goodbreak
\and
National University of Ireland, Department of Experimental Physics, Maynooth, Co. Kildare, Ireland\goodbreak
\and
Nicolaus Copernicus Astronomical Center, Bartycka 18, 00-716 Warsaw, Poland\goodbreak
\and
Niels Bohr Institute, Blegdamsvej 17, Copenhagen, Denmark\goodbreak
\and
Optical Science Laboratory, University College London, Gower Street, London, U.K.\goodbreak
\and
SISSA, Astrophysics Sector, via Bonomea 265, 34136, Trieste, Italy\goodbreak
\and
School of Physics and Astronomy, Cardiff University, Queens Buildings, The Parade, Cardiff, CF24 3AA, U.K.\goodbreak
\and
Sorbonne Universit\'{e}-UPMC, UMR7095, Institut d'Astrophysique de Paris, 98 bis Boulevard Arago, F-75014, Paris, France\goodbreak
\and
Space Research Institute (IKI), Russian Academy of Sciences, Profsoyuznaya Str, 84/32, Moscow, 117997, Russia\goodbreak
\and
Space Sciences Laboratory, University of California, Berkeley, California, U.S.A.\goodbreak
\and
Special Astrophysical Observatory, Russian Academy of Sciences, Nizhnij Arkhyz, Zelenchukskiy region, Karachai-Cherkessian Republic, 369167, Russia\goodbreak
\and
Sub-Department of Astrophysics, University of Oxford, Keble Road, Oxford OX1 3RH, U.K.\goodbreak
\and
UPMC Univ Paris 06, UMR7095, 98 bis Boulevard Arago, F-75014, Paris, France\goodbreak
\and
Universit\'{e} de Toulouse, UPS-OMP, IRAP, F-31028 Toulouse cedex 4, France\goodbreak
\and
University of Granada, Departamento de F\'{\i}sica Te\'{o}rica y del Cosmos, Facultad de Ciencias, Granada, Spain\goodbreak
\and
University of Granada, Instituto Carlos I de F\'{\i}sica Te\'{o}rica y Computacional, Granada, Spain\goodbreak
% ...
}

\abstract{We use \Planck\ data to detect the cross-correlation
          between the thermal Sunyaev-Zeldovich (tSZ) effect and the infrared
          emission from the galaxies that make up the the cosmic infrared
          background (CIB).  We first
          perform a stacking analysis towards \Planck-confirmed
          galaxy clusters.  We detect infrared emission produced by dusty
          galaxies inside these clusters and demonstrate that the
          infrared emission is about 50\,\% more extended than the tSZ effect.
          Modelling the emission with a Navarro--Frenk--White profile, we find
          that the radial profile concentration parameter is $c_{500} =
          1.00^{+0.18}_{-0.15}$.  This indicates that infrared galaxies in the
          outskirts of clusters have higher infrared flux than cluster-core
          galaxies.  We also study the cross-correlation between tSZ and CIB
          anisotropies, following three alternative approaches based on
          power spectrum analyses: (i) using a catalogue of confirmed
          clusters detected in \Planck\ data; (ii) using an all-sky tSZ map
          built from \Planck\ frequency maps; and
          (iii) using cross-spectra between \Planck\ frequency maps.
          With the three different methods, we detect
          the tSZ-CIB cross-power spectrum at significance levels of
          (i) 6\,$\sigma$, (ii) 3\,$\sigma$, and (iii) 4\,$\sigma$.  We
          model the tSZ-CIB cross-correlation signature and compare
          predictions with the measurements.  The 
          amplitude of the cross-correlation relative to the fiducial
          model is $A_{\rm tSZ-CIB}= 1.2\pm0.3$.  This
          result is consistent with predictions for
          the tSZ-CIB cross-correlation assuming the best-fit cosmological
          model from \Planck\ 2015 results along with the tSZ and CIB scaling
          relations.}

\authorrunning{Planck Collaboration}
\titlerunning{The tSZ--CIB cross-correlation}

\maketitle 

 
\section{Introduction}

This paper is one of a set associated with the 2015 release of data
from the \Planck\footnote{\Planck\ (\url{http://www.esa.int/Planck})
 is a project of the 
 European Space Agency  (ESA) with instruments provided by two scientific 
 consortia funded by ESA member states and led by Principal Investigators 
 from France and Italy, telescope reflectors provided through a collaboration 
 between ESA and a scientific consortium led and funded by Denmark, and 
 additional contributions from NASA (USA).}
mission. It reports the first all-sky detection of the
cross-correlation between the thermal Sunyaev-Zeldovich (tSZ) effect
\citep{sun69,sun72} and the cosmic infrared background
\citep[CIB;][]{pug96,fix98,hau98}.  An increasing number of
observational studies are measuring the tSZ effect and CIB fluctuations at
infrared and submillimetre wavelengths, including
investigations of the CIB 
with the {\it Spitzer Space Telescope} \citep{lag07} and the
{\it Herschel Space Observatory\/} \citep{amb11,vie12,vie15}, and
observations of the tSZ effect with instruments such as the Atacama
Pathfinder Experiment \citep{hal09} and Bolocam \citep{say11}.  In
addition, a new generation of CMB experiments can measure the tSZ
effect and CIB at microwave frequencies
\citep{hin10,hal10,dun11,zwa11,rei12,planck2013-p05b,planck2013-pip56}.

The large frequency coverage of \Planck, from 30 to 857\,GHz,
makes it sensitive to both of these important probes of large-scale
structure.  At intermediate frequencies,
from 70 to 217\,GHz, the sky emission is dominated by the cosmic
microwave background (CMB). At these frequencies, it is possible to
detect galaxy clusters that produce a distortion of the CMB blackbody
emission through the tSZ effect. At the angular resolution of \Planck,
this effect is mainly produced by local ($z < 1$) and massive galaxy
clusters in dark matter halos (above $10^{14}\,{\rm M_\odot}$), and it
has been used for several studies of cluster physics and cosmology
\citep[e.g.,][]{planck2011-5.2a,planck2011-5.2b,planck2012-III,planck2012-V,
planck2012-VIII,planck2012-X,planck2013-p15,planck2014-a30,planck2013-p15,
planck2014-a30}.
At frequencies above 353\,GHz,
the sky emission is dominated by thermal emission, both Galactic and
extragalactic \citep{planck2013-p06b,planck2013-pip56}.
The dominant extragalactic signal is the thermal infrared emission from
dust heated by UV radiation from
young stars. According to our current knowledge of star-formation history,
the CIB emission has a peak in its redshift distribution between
$z=1$ and $z=2$, and is produced by galaxies in dark matter halos
of $10^{11}$--$10^{13}\,{\rm M_\odot}$; this has been confirmed through the
measured strong
correlation between the CIB and CMB lensing \citep{planck2013-p13}.
However, due to the different redshift and mass ranges, the CIB and tSZ
distributions have little overlap at the angular scales probed by
\Planck, making this correlation hard to detect. 

Nevertheless, determining the strength of this tSZ-CIB correlation is
important for several reasons.  Certainly we need to know the extent to which
tSZ estimates might be contaminated by CIB fluctuations,
but uncertainty in the
correlation also degrades our ability to estimate power coming from the
``kinetic'' SZ effect (arising from peculiar motions), which promises to
probe the reionization epoch \citep[e.g.,][]{mes12,rei12,planck2015-XXXVII}.
But as well as this, analysis of the tSZ-CIB correlation
allows  us to better understand the spatial distribution and evolution of
star formation within massive halos.

The profile of infrared emission from galaxy clusters is expected to be less
concentrated than the profile of the number counts of galaxies. Indeed, core
galaxies present reduced infrared emission owing to quenching, which 
occurs after they make their first passage inside $r \simeq R_{500}$
\citep{muz14}.  Using SDSS data, \citet{wei10} computed the radial
profile of passive galaxies for high-mass galaxy clusters
($M > 10^{14}\,{\rm M_\odot}$). They found that the fraction of passive
galaxies is 70--80\,\% at the centres and 
25--35\,\% in the outskirts of clusters.  The detection of infrared emission
cluster by cluster is difficult at millimetre wavelengths, since the emission
is faint and confused by the fluctuations of the infrared sky (Galactic
thermal dust and CIB). Statistical
detections of infrared emission in galaxy clusters have been
made by stacking large samples of known clusters
\citep{mon05,gia08,ron10} in IRAS data \citep{whe93}.
The stacking approach has also been shown to be a powerful method for extracting
the tSZ signal from microwave data \citep[e.g.,][]{lie06,die09}.

Recently, efforts have been made to model the
tSZ-CIB correlation \citep[e.g.,][]{zah12,add12}.  Using a halo model,
it is possible to predict the tSZ-CIB cross-correlation. The
halo model approach allows us to consider distinct astrophysical
emission processes that trace the large-scale dark matter density fluctuation,
but have different dependencies on the host halo mass and
redshift. In this paper, we use models of the tSZ-CIB cross-correlation
at galaxy cluster scales. We note that the tSZ effect does not possess
significant substructure on the scale of galaxies, so
the tSZ-CIB cross-correlation should not possess a shot noise term.

Current experiments have already provided constraints on
the tSZ-CIB cross-correlation at low frequencies, between 100 and 250\,GHz.
The ACT collaboration set an upper limit $\rho < 0.2$ on the tSZ-CIB cross
correlation \citep{dun13}.  \citet{geo14}, using SPT data and  assuming a  single correlation factor, obtained a
tSZ-CIB correlation factor of $0.11^{+0.06}_{-0.05}$;
a zero correlation is disfavoured at a confidence level of 99\,\%.

Our objective in this paper is twofold. First, we characterize the CIB
emission toward tSZ-detected galaxy clusters by constraining the profile and
redshift dependence of CIB emission from galaxy clusters. Then, we set
constraints on the overall tSZ-CIB cross-correlation power spectrum, and
report the first all-sky detection of the tSZ-CIB angular cross-power spectra,
at a significance level of $4\,\sigma$.
Our models and results on the tSZ-CIB cross-correlation 
have been used in a companion \Planck\ paper \cite{planck2014-a28}. 

In the first part of Sect.~\ref{secmod1}, we explain our modelling approach
for the tSZ effect and CIB emission at the galaxy cluster scale. 
In the second part of Sect.~\ref{secmod1}, we describe the model for the tSZ,
CIB, and tSZ-CIB power and cross-power spectra using a halo model.
Then in Sect.~\ref{secdat} we present the data sets we have used.
Sections~\ref{secres1} and~\ref{secres2}
present our results for the  SED, shape, and cross-spectrum of
the tSZ-CIB correlation.  Finally, in Sect.~\ref{seccon} we discuss
the results and their consistency with previous analyses.

Throughout this paper, the tSZ effect intensity will be expressed in units
of Compton parameter, and we will use the best-fit cosmology from
\citet[][fourth column of table~3]{planck2014-a15} using
``TT,TE,EE+lowP'' values as the fiducial
cosmological model, unless otherwise specified. Thus, we adopt $H_0
= 67.27\,{\rm km}\,{\rm s}^{-1}\,{\rm Mpc}^{\-1}$, $\sigma_8 = 0.831$,
and $\Omega_{\rm m} = 0.3156$.


\section{Modelling}
\label{secmod1} 

To model the cross-correlation between tSZ and CIB anisotropies
we have to relate the
mass, $M_{500}$, and the redshift, $z$, of a given cluster to tSZ flux,
$Y_{500}$, and CIB luminosity $L_{500}$. We define $M_{500}$ (and $R_{500}$)
as the total mass (and radius) for which the mean over-density 
is 500 times the critical density of the Universe.
Considering that the tSZ signal in the \Planck\ data has no
significant substructure at galaxy scales, we modelled
the tSZ-CIB cross-correlation at the galaxy cluster
scale. This can be considered as a large-scale approximation for the
CIB emission, and at the \Planck\ angular
resolution it agrees with the more refined modelling presented in
\citet{planck2013-pip56}.

\begin{table}[!tbh]
\begingroup
\newdimen\tblskip \tblskip=5pt
\caption{Cosmological and scaling-law parameters for our fiducial model, for
both the $Y_{500}$--$M_{500}$ relation \citep{planck2013-p15}
and the $L_{500}$--$M_{500}$
relation (fitted to spectra from \citealt{planck2013-pip56}).}
\label{tabscal}
\nointerlineskip
\vskip -3mm
\footnotesize
\setbox\tablebox=\vbox{
 \newdimen\digitwidth
 \setbox0=\hbox{\rm 0}
  \digitwidth=\wd0
  \catcode`*=\active
  \def*{\kern\digitwidth}
  \newdimen\signwidth
  \setbox0=\hbox{+}
  \signwidth=\wd0
  \catcode`!=\active
  \def!{\kern\signwidth}
\halign{\tabskip=0pt\hbox to 2.0cm{#\leaderfil}\tabskip=0.5em&
  \hfil#\hfil\tabskip=0pt\cr
\noalign{\doubleline}
\multispan2\hfil \Planck-SZ cosmology\hfil\cr
$\Omega_{\rm m}$&     $*0.29\pm0.02$\cr
$\sigma_8$&           $*0.77\pm0.02$\cr
$H_0$&                $67.3\pm1.4$\cr
\noalign{\vskip 4pt\hrule\vskip 5pt}
\multispan2\hfil \Planck-CMB cosmology\hfil\cr
$\Omega_{\rm m}$&     $*0.316\pm0.009$\cr
$\sigma_8$&           $*0.831\pm0.013$\cr
$H_0$&                $67.27\pm 0.66$\cr
\noalign{\vskip 4pt\hrule\vskip 5pt}
\multispan2\hfil $M_{500}$--$Y_{500}$\hfil\cr
${\rm log}\,Y_\ast$&  $-0.19\pm0.02$\cr
$\alpha_{\rm SZ}$&    $!1.79\pm0.08$\cr
$\beta_{\rm SZ}$&     $!0.66\pm0.50$\cr
\noalign{\vskip 4pt\hrule\vskip 5pt}
\multispan2\hfil $M_{500}$--$L_{500}$\hfil\cr
$T_0$&                $24.4\pm1.9$\cr
$\alpha_{\rm CIB}$&   $*0.36\pm0.05$\cr
$\beta_{\rm CIB}$&    $*1.75\pm0.06$\cr
$\gamma_{\rm CIB}$&   $*1.70\pm0.02$\cr
$\delta_{\rm CIB}$&   $*3.2*\pm0.2*$\cr
$\epsilon_{\rm CIB}$& $*1.0*\pm0.1*$\cr
\noalign{\vskip 4pt\hrule\vskip 3pt}}}
\endPlancktable
\endgroup
\end{table}

\subsection{The thermal Sunyaev-Zeldovich effect}

The tSZ effect is a small-amplitude distortion of the CMB
blackbody spectrum caused by inverse-Compton scattering
\citep[see, e.g.,][]{Rephaeli1995,Birkinshaw1999,Carlstrom2002}.  Its intensity
is related to the integral of the pressure along the line of sight via the
Compton parameter, which for a given direction on the sky is
\begin{equation}
y = \int \frac{k_{\rm B} \sigma_{\rm T}}{m_{\rm e} c^2} n_{\rm e}
 T_{\rm e} {\rm d}l.
\end{equation}
Here ${\rm d}l$ is the distance along the line of sight,
$k_{\rm B}$, $\sigma_{\rm T}$, $m_{\rm e}$, and $c$ are the usual physical
constants, and $n_{\rm e}$ and $T_{\rm e}$ are the electron number density
and the temperature, respectively.

In units of CMB temperature, the contribution of the tSZ effect to the
submillimetre sky intensity for a given observation frequency $\nu$ is given by
\begin{equation}
\frac{\Delta T_{\rm CMB}}{T_{\rm CMB}} = g(\nu) y.
\end{equation}
Neglecting relativistic corrections we have $g(\nu) = x \coth(x/2) - 4$, with $x = h\nu/(k_{\rm B} T_{\rm CMB})$.
The function $g(\nu)$ is is equal to 0 at about 217\,GHz, and is negative at
lower frequencies and positive at higher frequencies.

We have used the $M_{500}$--$Y_{500}$ scaling law presented in
\citet{planck2013-p15}, 
\begin{equation}
E^{-\beta_{\rm SZ}}(z) \left[ \frac{D^2_{\rm A}(z)
    {Y}_{500}}{10^{-4}\,{\rm Mpc}^2} \right] = Y_\ast \left[
  \frac{h}{0.7} \right]^{-2+\alpha_{\rm SZ}} \left[ \frac{(1-b)
    M_{500}}{6 \times 10^{14}\,{\rm M_{\odot}}}\right]^{\alpha_{\rm SZ}},
\label{szlaw}
\end{equation}
with $E(z) = \sqrt{\Omega_{\rm m}(1+z)^3 + \Omega_{\Lambda}}$ for a flat
universe. The coefficients $Y_\ast$, $\alpha_{\rm SZ}$, and $\beta_{\rm SZ}$
are taken from \citet{planck2013-p15}, and are given in Table~\ref{tabscal}.
The mean bias, $(1-b)$, between X-ray mass and the true mass is discussed
in detail in \citet[appendix~A]{planck2013-p15} and references therein.
We adopt $b=0.3$ here, which, given the chosen cosmological parameters,
allows us to reproduce the tSZ results from \citet{planck2014-a28} and
\citet{planck2014-a30}.

\subsection{Cosmic infrared background emission}

The CIB is the diffuse emission from galaxies integrated
throughout cosmic history \citep[see, e.g.,][]{Hauser2001,Lagache2005},
and is thus strongly related to the star-formation rate history.
The CIB intensity, $I(\nu)$, at frequency $\nu$ can be written as
\begin{equation}
I(\nu) = \int {\rm d}z \frac{{\rm d}\chi(z)}{{\rm d}z}\,
 \frac{j(\nu,z)}{(1+z)},
\end{equation}
with $\chi(z)$ the comoving distance and $j(\nu,z)$ the emissivity that
is related to the star-formation density, $\rho_{\rm SFR}$, through
\begin{equation}
 j(\nu,z) = \frac{\rho_{\rm SFR}(z) (1+z) \Theta_{\rm eff}(\nu,z) \chi^2(z)}{K},
\end{equation}
where $K$ is the \citet{ken98} constant (SFR/$L_{\rm IR} = 1.7\times
10^{-10}\,{\rm M_{\odot}}\,{\rm yr}^{-1}$) and $\Theta_{\rm eff}(\nu,z)$
the mean spectral energy distribution (SED) of infrared galaxies at redshift
$z$.

To model the $L_{500}$--$M_{500}$ relation we use a parametric
relation proposed by \citet{sha12} that relates the CIB flux,
$L_{500}$, to the mass, $M_{500}$, as follows:
\begin{equation}
L_{500}(\nu) = L_0 \left[ \frac{M_{500}}{1 \times 10^{14}\,{\rm
      M_{\odot}}} \right]^{\epsilon_{\rm CIB}}\Psi(z) \, \Theta \left[
  (1+z)\nu,T_{\rm d}(z) \right],
\label{ciblm}
\end{equation}
where $L_0$ is a normalization parameter, $T_{\rm d}(z) = T_{{\rm d}0}
(1+z)^{\alpha_{\rm CIB}}$ and $\Theta \left[ \nu,T_{\rm d}\right]$ is
the typical SED of a galaxy that contributes to the total CIB
emission,
$$
\Theta \left[ \nu,T_{\rm d}\right] = \left\{
    \begin{array}{ll}
      \nu^{\,\beta_{\rm CIB}} B_{\nu}(T_{\rm d}), & \mbox{if} \ \nu < \nu_0,\\
      \nu^{-\gamma_{\rm CIB}}, & \mbox{if} \  \nu \geq \nu_0,
    \end{array}
\right.
$$ 
with $\nu_0$ being the solution of ${\rm d\, log}
 [\nu^{\,\beta_{\rm CIB}} B_{\nu}(T_{\rm d})]/
 {\rm d\,log}(\nu) = -\gamma_{\rm CIB}$.  We assume a redshift dependence
of the form
\begin{equation}
\Psi(z) = (1+z)^{\delta_{\rm CIB}}.
\end{equation}
We also define $S_{500}$ as
\begin{equation}
\label{eqs500}
{S}_{500}(\nu) = \frac{L_{500}(\nu)}{4 \pi (1+z)\chi^2(z)}.
\end{equation}

The coefficients $T_{{\rm d}0}$, $\alpha_{\rm CIB}$, $\beta_{\rm CIB}$,
$\gamma_{\rm CIB}$, and $\delta_{\rm CIB}$ from \citet{planck2013-pip56} are
given in Table~\ref{tabscal}. We fix the value of $\epsilon_{\rm CIB}$ to 1. In Sect.~\ref{secauto}, this model of the CIB
emission will be compared with the \Planck\ measurement of the CIB power spectra.
We stress that this parametrization can only be considered accurate at
scales where galaxy clusters are not (or only marginally) extended.
This is typically the case at \Planck\ angular resolution for the low-mass and
high-redshift dark matter halos that dominate the total CIB emission.


\subsection{Angular power spectra}
\label{secmod2}
\subsubsection{The halo model}

\begin{figure}[!th]
\begin{center}
\includegraphics[scale=0.20]{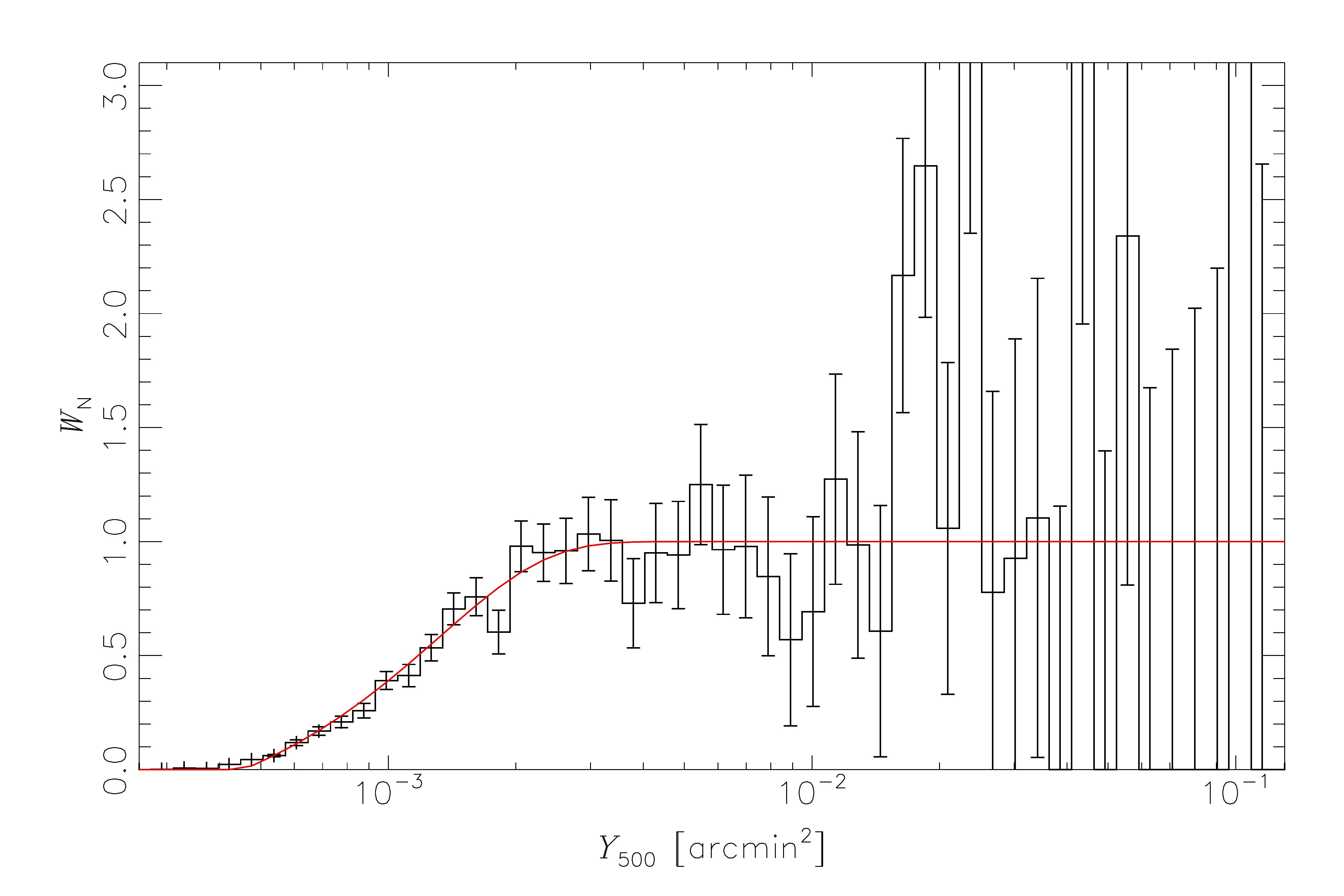}
\caption{Weight of the mass function as a function of $Y_{500}$. \textit{Black}: the ratio between the number of observed clusters
  and the predicted number of clusters from the \Planck-SZ best-fit
  cosmology. \textit{Red}: parametric formula, Eq.~(\ref{seleq}),
  for the selection function.}
\label{selfunc}
\end{center}
\end{figure}

\begin{figure*}[!th]
\begin{center}
\includegraphics[scale=0.4]{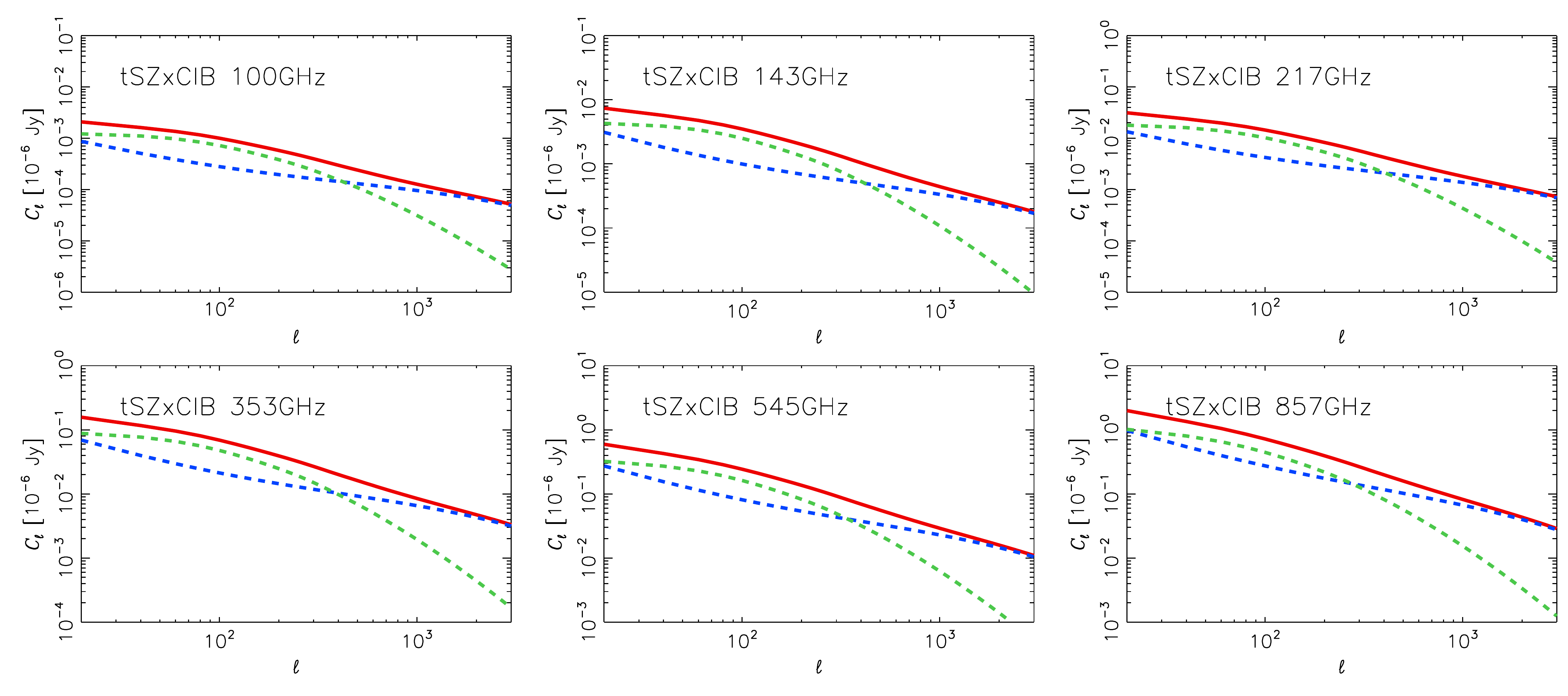}
\caption{Predicted tSZ-CIB cross-correlation from 100 to 857\,GHz for the
  fiducial model, where the tSZ signal is expressed in Compton parameter units.
  The blue dashed line presents the prediction for the 1-halo term, the
  green dashed line for the 2-halo term and the red solid line for the
  total model.}
\label{szcibspec}
\end{center}
\end{figure*}

To model tSZ, CIB, and tSZ-CIB angular power spectra, we consider the
halo-model formalism \citep[see, e.g.][]{cooray2002}
and the following general expression
\begin{equation}
C_{\ell} = C^{AB,{\rm 1h}}_\ell + C^{AB,{\rm 2h}}_\ell,
\end{equation}
where $A$ and $B$ stand for tSZ effect or CIB emission, $C^{AB, {\rm 1h}}_\ell$
is the 1-halo contribution, and $C^{AB,{\rm 2h}}_\ell$
is the 2-halo term that accounts for correlation in the spatial
distribution of halos over the sky.

The 1-halo term $C^{AB,{\rm 1h}}_\ell$ is computed
using the Fourier transform of the projected profiles of signals $A$
and $B$ weighted by the mass function and the $A$ and $B$ emission
\citep[see, e.g.,][for a derivation of the tSZ angular power spectrum]{kom02}:
\begin{equation}
C_{\ell}^{AB,{\rm 1h}} = 4 \pi \int {\rm d}z \frac{{\rm d}V}{{\rm
    d}z {\rm d}\Omega}\int{\rm d}M \frac{{\rm d^2N}}{{\rm d}M {\rm
    d}V} W^{A,{\rm 1h}} W^{B,{\rm 1h}} ,
\end{equation}
where ${\rm d}^2N/{\rm d}M{\rm d}V$ is the dark-matter halo
mass function from \citet{tin08}, ${\rm d}V/{\rm d}z {\rm d}\Omega$ is
the comoving volume element, and $W^{A,{\rm 1h}}$, $W^{B,{\rm 1h}}$ are
the window functions that account for selection effects and total halo signal.
For the tSZ effect, we have $W^{\rm 1h}_{\rm tSZ} = W_{\rm N}(Y_{500})
{Y}_{500} y_{\ell}(M_{500}, z)$ and $W_{\rm N}(Y_{500})$ is a weight,
ranging from 0 to 1, applied to the mass function to account for the
effective number of clusters used in our analysis; here ${Y}_{500}$ is the
tSZ flux, and $y_\ell$ is the Fourier transform of the tSZ profile.
For the CIB emission we have $W^{\rm 1h}_{\rm CIB} = {S}_{500}(\nu)
I_{\ell}(M_{500}, z)$, where $I_{\ell}$ the Fourier transform of the
infrared profile (from Eq.~\ref{cibpro}) and ${S}_{500}(\nu)$ is given
in Eq.~(\ref{eqs500}).

The results for the radial analysis in Sect.~\ref{seccompa} show that
the infrared emission profile can be well approximated by an NFW profile
\citep{nav97} with a concentration parameter $c_{500} = 1.0$.
The galaxy cluster pressure
profile is modelled by a generalized NFW profile \citep[GNFW,][]{nag07}
using the best-fit values from \citet{arn10}.

The contribution of the 2-halo term, $C^{AB,{\rm 2h}}_\ell$,
accounts for large-scale fluctuations in the matter power spectrum
that induce correlations in the dark-matter halo distribution over the
sky.  It can be computed as
\begin{align}
C_{\ell}^{AB,{\rm 2h}} = 4 \pi &\int {\rm d}z
 \frac{{\rm d}V}{{\rm d}z{\rm d}\Omega} \left(\int{\rm d}M
 \frac{{\rm d^2N}}{{\rm d}M {\rm d}V} W^{A,{\rm 1h}}
 b_{\rm lin}(M,z)\right) \nonumber \\
 &\times\left(\int{\rm d}M
 \frac{{\rm d^2N}}{{\rm d}M {\rm d}V} W^{B,{\rm 1h}}
 b_{\rm lin}(M,z)\right) P(k,z)
\end{align}
 \citep[see, e.g.,][and references therein]{tab11},
where $P(k,z)$ is the matter power spectrum computed using
{\tt CLASS} \citep{les11} and $b_{\rm lin}(M,z)$ is the time-dependent
linear bias factor \citep[see][for an analytical formula]{mo96,kom99,tin10b}.

As already stated, at the \Planck\ resolution the tSZ emission does not
have substructures at galaxy scales. Consequently there is no shot-noise in
the tSZ auto-spectrum and all the tSZ-related cross-spectra.  
Considering the method we used to compute the CIB auto-correlation power
spectra, the total amplitude of the 1-halo term should include this
``shot-noise'' (galaxy auto-correlation). We have verified that, at the resolution
of \Planck, there is no significant difference between our modelling and a
direct computation of the shot-noise using the sub-halo mass function from
\citet{tin10}, by comparing our modelling
with \Planck\ measurements of the CIB auto-spectra.

\subsubsection{Weighted mass-function for selected tSZ sample}

Some of our analyses are based on a  sample selected from a tSZ catalogue. 
However, we only consider confirmed galaxy clusters with known redshifts.
Therefore, it is not possible to use the selection function of the catalogue,
which includes some unconfirmed clusters. To account for our selection, we introduce a weight function,
$W_{\rm N}$, which we estimate by computing the tSZ flux $Y_{500}$ as a
function of the mass and redshift of the clusters through the $Y$--$M$
scaling relation.  Then we compute the ratio between the number of clusters
in our selected sample and the predicted
number (derived from the mass function) as a function of the flux
$Y_{500}$. We convolve the mass function with the scatter of the $Y$--$M$
scaling relation, to express observed and predicted quantities in a
comparable form.
Uncertainties are obtained assuming a Poissonian number count for the clusters
in each bin. Finally, we approximate this ratio with a parametric formula:
\begin{equation} 
W_{\rm N} = {\rm erf}(660\, Y_{500} - 0.30).
\label{seleq}
\end{equation}
This formula is a good approximation for detected galaxy clusters, with
$Y_{500} > 10^{-4}$ arcmin$^2$.
The weight function is presented in Fig.~\ref{selfunc}.  
This weight applied to the mass function is degenerate with the
cosmological parameters, and thus cancels cosmological parameter
dependencies of the tSZ-CIB cross-correlation.

Moreover, considering that the detection methods depend on both $Y_{500}$ and
$\theta_{500}$, a more accurate weight function could be defined in the
$Y_{500}$--$\theta_{500}$ plane and convolved with the variation of noise
amplitude across the sky.  However, given the low number of clusters
in our sample, we choose to define our weight only with respect to $Y_{500}$.

\begin{figure}[!ht]
\begin{center}
\includegraphics[scale=0.2]{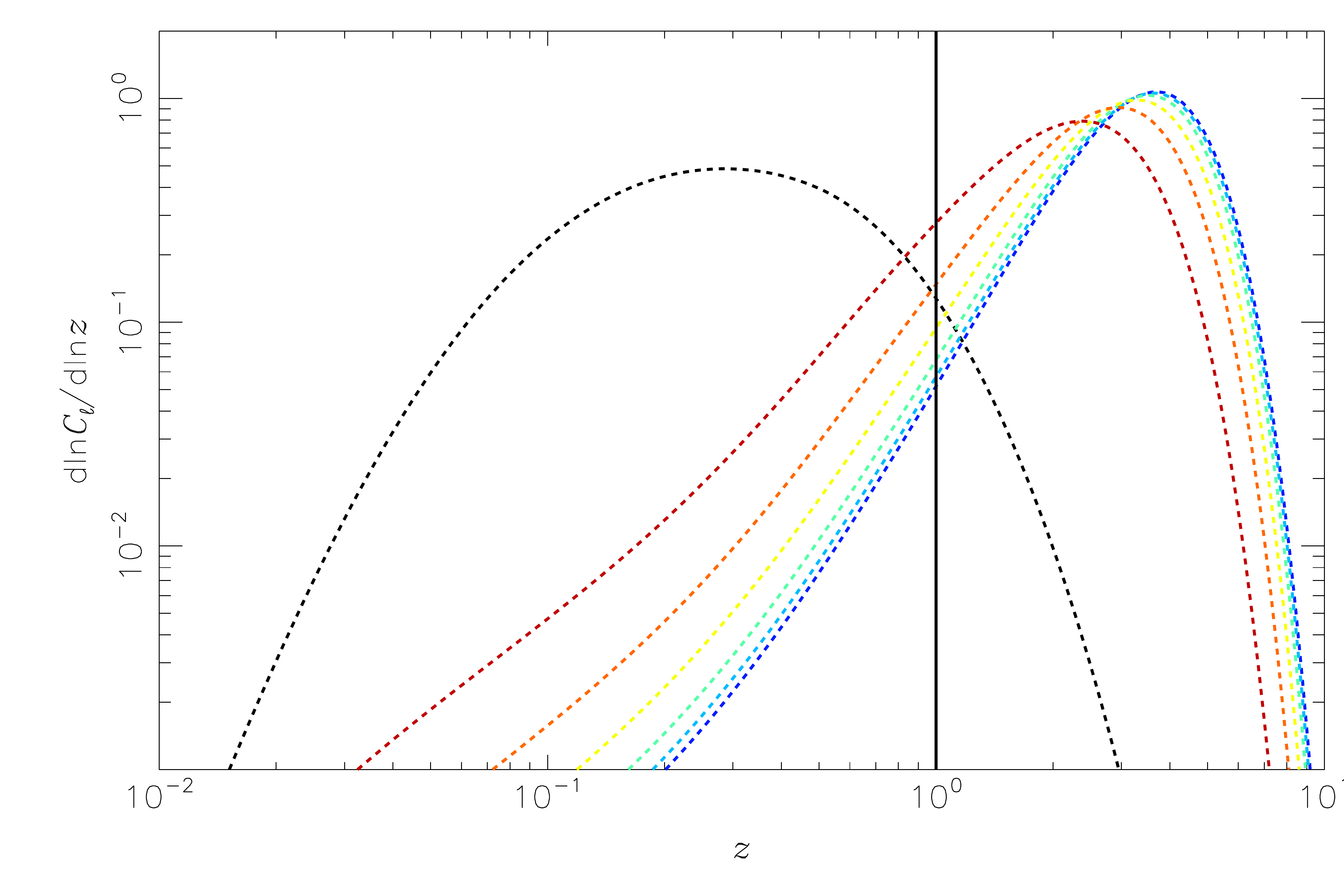}
\includegraphics[scale=0.2]{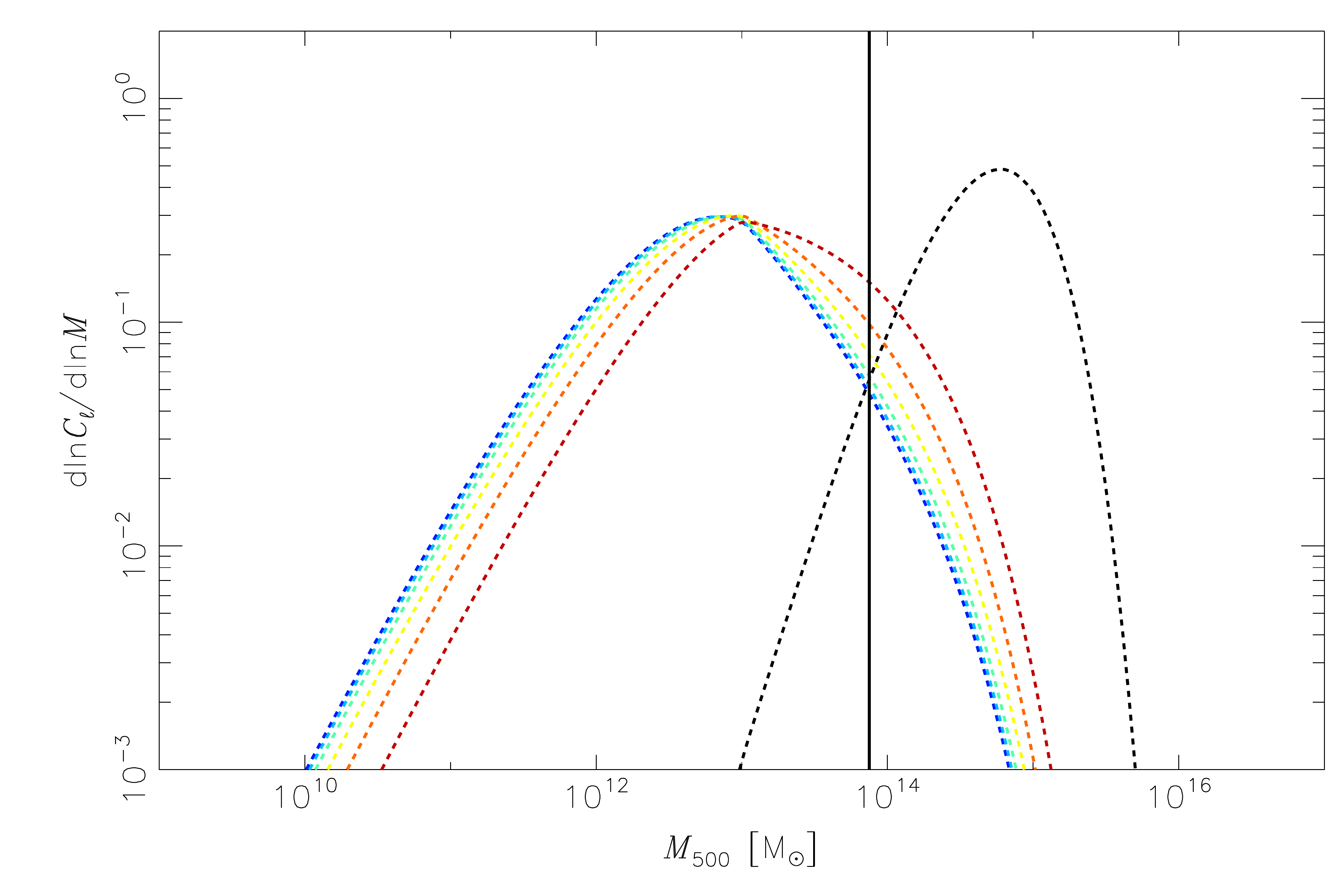}
\caption{{\it Top}: predicted distribution of the tSZ and CIB power as a
  function of the redshift at $\ell = 1000$. {\it Bottom}: predicted
  distribution of the tSZ and CIB power as a function of the host halo
  mass at $\ell = 1000$. The black dashed line is for the tSZ effect, while
  the dark blue, light blue, green, yellow, orange, and red dashed lines
  are for CIB at 100, 143, 217, 353, 545, and 857\,GHz
  respectively. The vertical solid black line shows the maximum
  redshift in PSZ2 (top panel) and the minimal $M_{500}$ in PSZ
  (bottom panel).}
\label{szcibdis}
\end{center}
\end{figure}

\begin{figure}[!th]
\begin{center}
\includegraphics[scale=0.2]{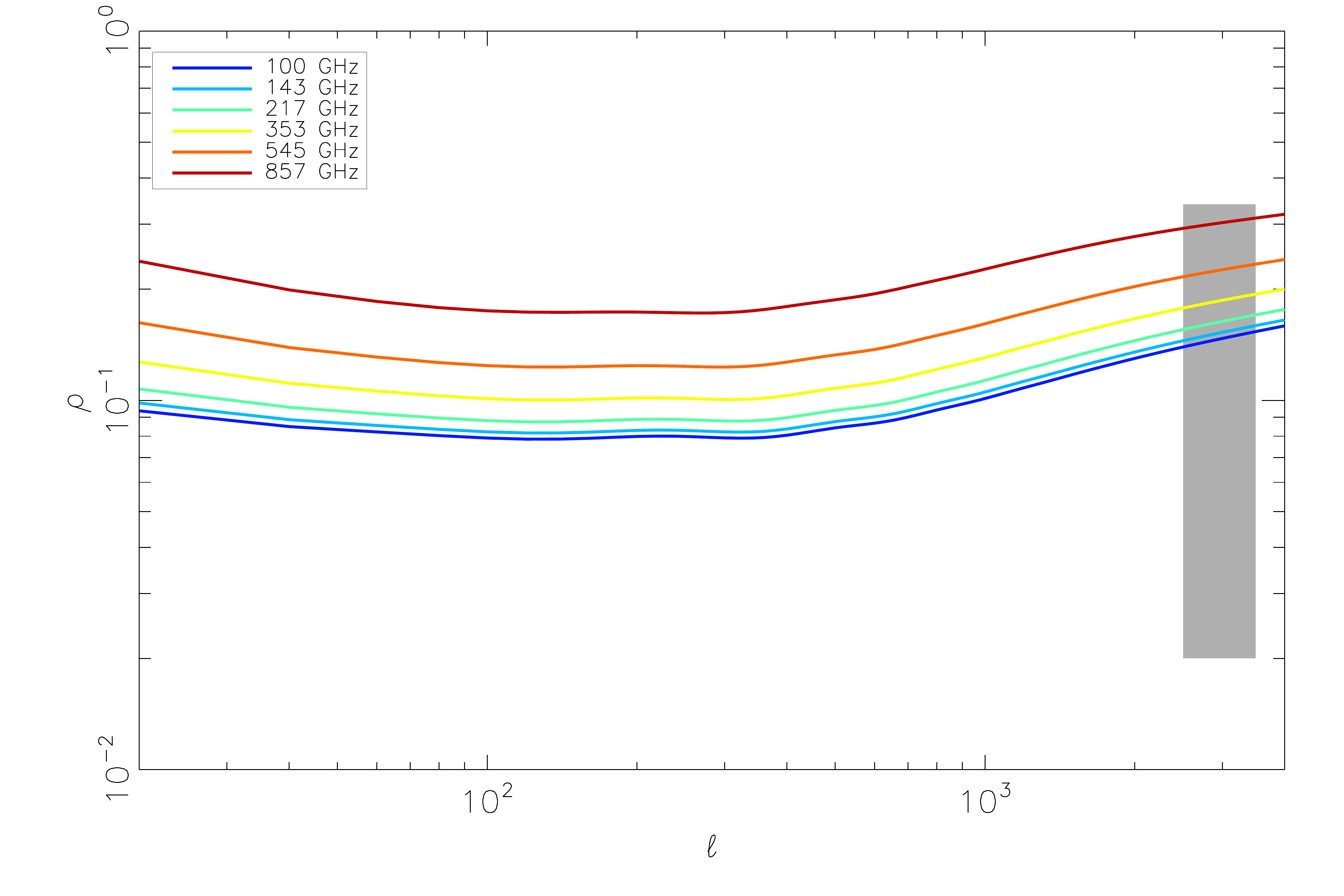}
\caption{Predicted correlation factor of the tSZ-CIB cross-spectrum from 100 to
  857\,GHz. The grey shaded area represents the range of values
  predicted for $\rho$ from \citet{zah12} for various models
  at 95, 150, and 220\,GHz.}
\label{szcibrho}
\end{center}
\end{figure}

\subsection{Predicted tSZ-CIB angular cross-power spectrum}

In Fig.~\ref{szcibspec} we present the predicted tSZ-CIB
angular cross-power spectra from 100 to 857\,GHz, for the fiducial
cosmological model and scaling-relation parameters listed in
Table~\ref{tabscal}.  The tSZ angular auto-spectrum is dominated by
the 1-halo term, while the CIB auto-spectrum is dominated by the
2-halo-term up to $\ell \simeq 2000$. Thus we need to
consider both contributions for the total cross-power spectrum.

At low $\ell$, we observe that the 2-halo term has a similar
amplitude to the 1-halo term at all frequencies. The 1-halo term
completely dominates the total angular cross-power spectrum up
to $\ell \simeq 2000$.  We also notice that the cross-power spectrum is highly
sensitive to the parameters $\delta_{\rm CIB}$ and $\epsilon_{\rm CIB}$.
Indeed, these two parameters set the overlap of the tSZ and CIB window
functions in mass and redshift.  Similarly, the relative amplitude of the
1-halo and 2-halo term is directly set by these parameters.

In Fig.~\ref{szcibdis}, we present the redshift and mass distribution
of tSZ and CIB power. These distributions are different for different
multipoles. Given the angular scale probed by \Planck, we show them for the
specific multipole $\ell = 1000$.  We notice that the correlation between
tSZ and CIB, at a given frequency, is determined by the overlap of these
distributions.  Clusters that constitute the main contribution to the
total tSZ power are at low redshifts ($z < 1$).  The galaxies that
produce the CIB are at higher redshifts ($1< z < 4$).  The mass
distribution of CIB power peaks near $M_{500} = 10^{13}\,{\rm M_\odot }$,
while the tSZ effect is produced mostly by halos in the range
$10^{14}\,{\rm M_\odot} < M_{500} < 10^{15}\,{\rm M_\odot}$.  

\begin{figure*}[!th]
\begin{center}
\includegraphics[scale=0.25]{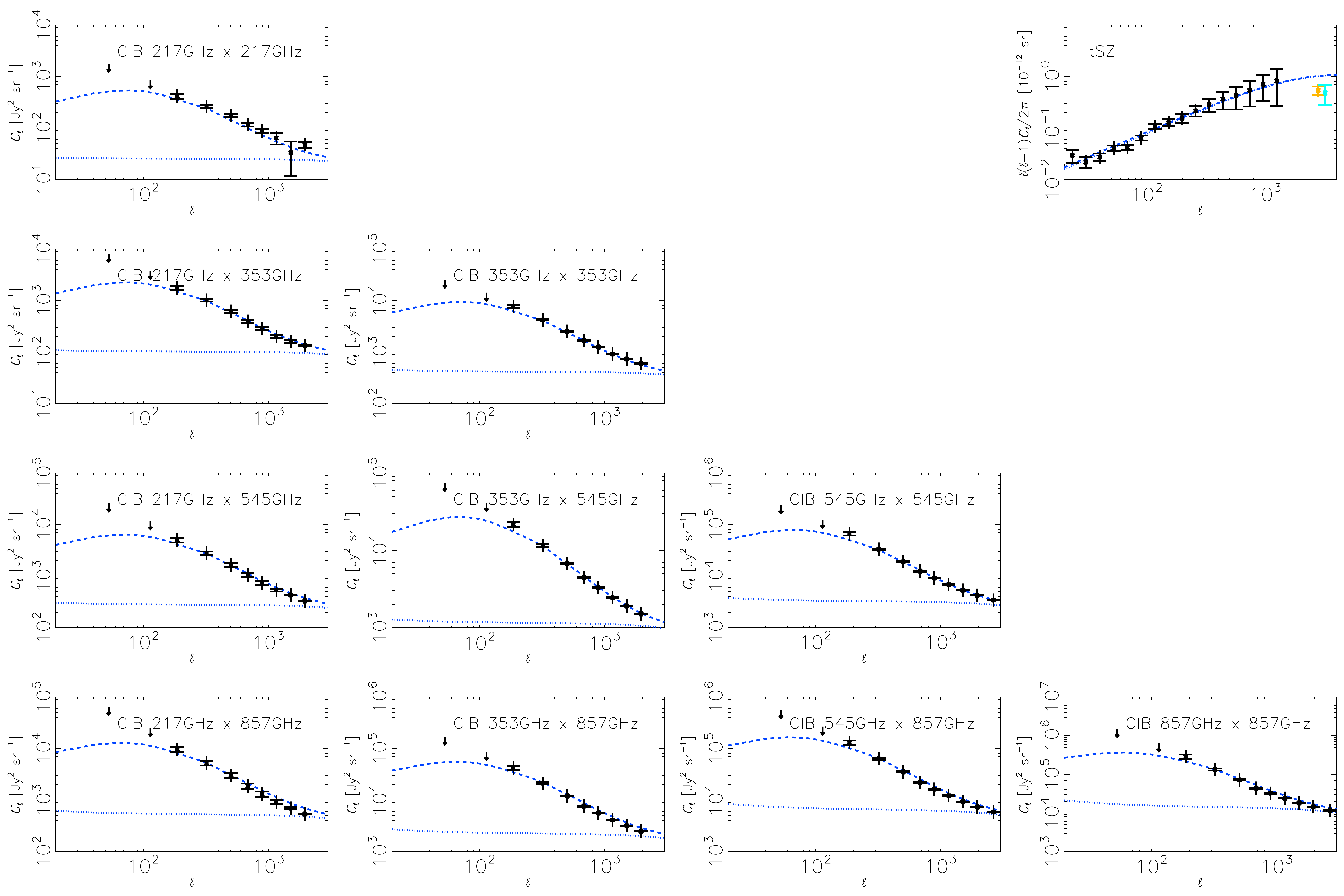}
\caption{\textit{Upper right panel}: observed tSZ power spectrum:
\Planck\ data from \citet{planck2013-p05b} (black symbols),
ACT data \citep{rei12} (light blue symbols), and SPT data \citep{sie13} (orange symbols); with
our fiducial model (dashed blue line). 
\textit{Other panels}:
observed CIB power spectra, with \Planck\ data from
\citet{planck2013-pip56}  (black) and our fiducial model
(dashed blue line). These panels show auto- and cross-power
spectra at 217, 353, 545, and 857\,GHz. The dotted blue lines show the 1-halo plus shot-noise
term for our fiducial model.}
\label{cibspec}
\end{center}
\end{figure*}

We define the correlation factor between tSZ and CIB signals, $\rho$, as
\begin{equation}
\rho = \frac{C_\ell^{{\rm tSZ-CIB}}}
 {\left(C_\ell^{\rm tSZ-tSZ}C_\ell^{\rm CIB-CIB}\right)^{1/2}}.
\end{equation}
Figure~\ref{szcibrho} shows that we
derive a correlation factor ranging from 0.05 to 0.30 at \Planck\ frequencies. This agrees
with the values reported from other tSZ-CIB modelling in the literature
\citep{zah12,add12}, ranging from 0.02 to 0.34 at 95, 150, and 220
GHz. The difference in redshift and mass distributions of tSZ and CIB
signals explains this relatively low degree of correlation.

We also observe that the tSZ-CIB correlation has a minimum around
$\ell=300$, and significantly increases at higher multipoles.  At those
multipoles, the tSZ effect is dominated by low-mass and higher-redshift
objects, overlapping better with the CIB range of masses and redshifts, which
explains the increase of the correlation factor. The frequency dependence of
the tSZ-CIB correlation factor can be explained by the variation of
the CIB window in redshift as a function of frequency.  At high
frequencies, we observe low-redshift objects
(with respect to other frequencies, but high-redshift objects from a
tSZ perspective). On the other hand, at low frequencies, we are sensitive to
higher redshift, as shown in Fig.~\ref{szcibdis}.

\subsection{Comparison with tSZ and CIB auto-spectra}
\label{secauto}

We fixed $T_{{\rm d}0}$, $\alpha_{\rm CIB}$, $\beta_{\rm CIB}$,
$\gamma_{\rm CIB}$, and $\delta_{\rm CIB}$ to the values from
\citet{planck2013-pip56}. We fix $\epsilon_{\rm CIB}$ to a value of
1.0, and we fit for $L_0$ in the multipole range $100 < \ell < 1000$
using CIB spectra from 217 to 857\,GHz.
We notice that the value of $\epsilon_{\rm CIB}$ is closely related to
halo occupation distribution power-law index and highly degenerate with
$\Omega_{\rm m}$.

In Fig.~\ref{cibspec}, we compare our modelling of the tSZ and CIB
spectra with measured spectra from the \Planck\ tSZ analysis
\citep{planck2014-a28} and \Planck\ data at 217, 353, 545, and 857\,GHz
\citep{planck2013-pip56}. For more details of these measurements see the related
\Planck\ papers referenced in \citet{planck2013-p01} and \citet{planck2014-a01}.
Here, we used the \Planck-SZ best-fit cosmology presented in
Table~\ref{tabscal}.
We observe that our model reproduces the observed auto-power spectra for
both tSZ and CIB anisotropies, except at low $\ell$ (below 100),
where the CIB power spectra are still contaminated by foreground Galactic
emission. For this reason, in this $\ell$ range the measured CIB power spectra
have to be considered as upper limits.  The figures show the consistency
of the present CIB modelling at cluster scale with the modelling
presented in \citet{planck2013-pip56} in the multipole range covered by the
\Planck\ data. Note that the flatness of the 1-halo term for CIB spectra
ensures that this term encompasses the shot-noise part.


\section{The data}
\label{secdat}

\subsection{\textit{Planck} frequency maps}
In this analysis, we use the \Planck\ full-mission data set
\citep{planck2014-a01,planck2014-a09}.
We consider intensity maps at frequencies from 30 to 857\,GHz, with
1\farcm7 pixels, to appropriately sample the resolution of the higher-frequency 
maps.  For the tSZ transmission in \Planck\
spectral bandpasses, we use the values provided in \citet{planck2013-p03d}.
We also used the bandpasses from \citet{planck2013-p03d} to compute the CIB
transmission in \Planck\ channels for the SED.
For power spectra analyses, we use \Planck\ beams from
\citet{planck2014-a05} and \citet{planck2013-p03c}.

\subsection{The \textit{Planck} SZ sample}
\label{secdats}
In order to extract the tSZ-CIB cross-correlation, we search for
infrared emission in the direction of clusters detected through their
tSZ signal.  In this analysis, we use galaxy clusters from the \Planck\ SZ
catalogue \citep[][PSZ2 hereafter]{planck2014-a36} that have measured
redshifts.  We restrict our analysis to the sample of confirmed clusters to
avoid contamination by false detections \citep[see][for more details]{agh14}.
This leads to a sample of 1093 galaxy clusters with a mean redshift
$\bar{z} \simeq 0.25$.

From this sample of clusters, we have built a reprojected tSZ
map.  We use a pressure profile from \citet{arn10}, with the scaling
relation presented in \citet{planck2013-p15}, as well as the size
($\theta_{500}$) and
flux ($Y_{500}$) computed from the 2-D posterior distributions
delivered
in \citet{planck2014-a36}.\footnote{\url{http://pla.esac.esa.int/pla/}}  We project each cluster onto an
oversampled grid with a pixel size of $0.1\times \theta_{500}$ (using
drizzling to avoid flux loss during the projection).  Then we convolve the
oversampled map with a beam of 10\arcm\ FWHM.  We reproject the oversampled
map onto a {\tt HEALPix} \citep{gorski2005} full-sky map with 1\farcm7 pixels
($N_{\rm side}=2048$) using nearest-neighbour interpolation.

\subsection{IRAS data}

We use the reprocessed IRAS maps, IRIS \citep[Improved Reprocessing of
the IRAS Survey,][]{miv05} in the {\tt HEALPix} pixelization scheme.
These offer improved calibration, zero level estimation, zodiacal light
subtraction, and destriping of the IRAS data. The IRIS 100, 60, and
25\,$\mu$m maps are used at their original resolution. Missing pixels in
the IRIS maps have been set to zero for this analysis.


\section{Results for tSZ-detected galaxy clusters.}
\label{secres1}

In this section, we present a detection of the tSZ-CIB cross-correlation
using known galaxy clusters detected via the tSZ effect in the \Planck\ data.
In Sect.~\ref{secstack}, we focus on the study of the
shape and the SED of the infrared emission towards galaxy clusters.
Then Sect.~\ref{seccat} is dedicated to the
study of the tSZ-CIB cross-power spectrum for confirmed tSZ clusters.

\subsection{Infrared emission from clusters}
\label{secstack}
\subsubsection{Stacking of \Planck\ frequency maps}

To increase the significance of the detection of infrared emission at
galaxy-cluster scales, we perform a stacking analysis of the sample of SZ
clusters defined in Sect.~\ref{secdat}. Following the methods presented in
\citet{hur14}, we extract individual patches of $4\deg\times4\deg$
from the full-sky \Planck\ intensity maps and
IRIS maps centred at the position of each cluster. The individual
patches are re-projected using a nearest-neighbour interpolation on a
grid of 0\farcm2 pixels in gnomonic projection to conserves the
cluster flux.  We then produce one stacked patch for each frequency. To do
so, the individual patches per frequency are co-added with a constant
weight. This choice accounts for the fact that the main contribution
to the noise, i.e., the CMB, is similar from one patch to 
another. Furthermore, it avoids cases where a particular cluster
dominates the stacked signal.  Considering that Galactic thermal dust
emission is not correlated with extragalactic objects, emission from
our Galaxy should not bias our stacking analysis. We verified that the
stacking is not sensitive to specific orientations, which may be
produced by the thermal dust emission from the Galactic plane. We
produced a stacked patch per frequency for the whole cluster sample,
and for two large redshift bins (below and above $z=0.15$).

\begin{figure}[!ht]
\begin{center}
\includegraphics[scale=0.17]{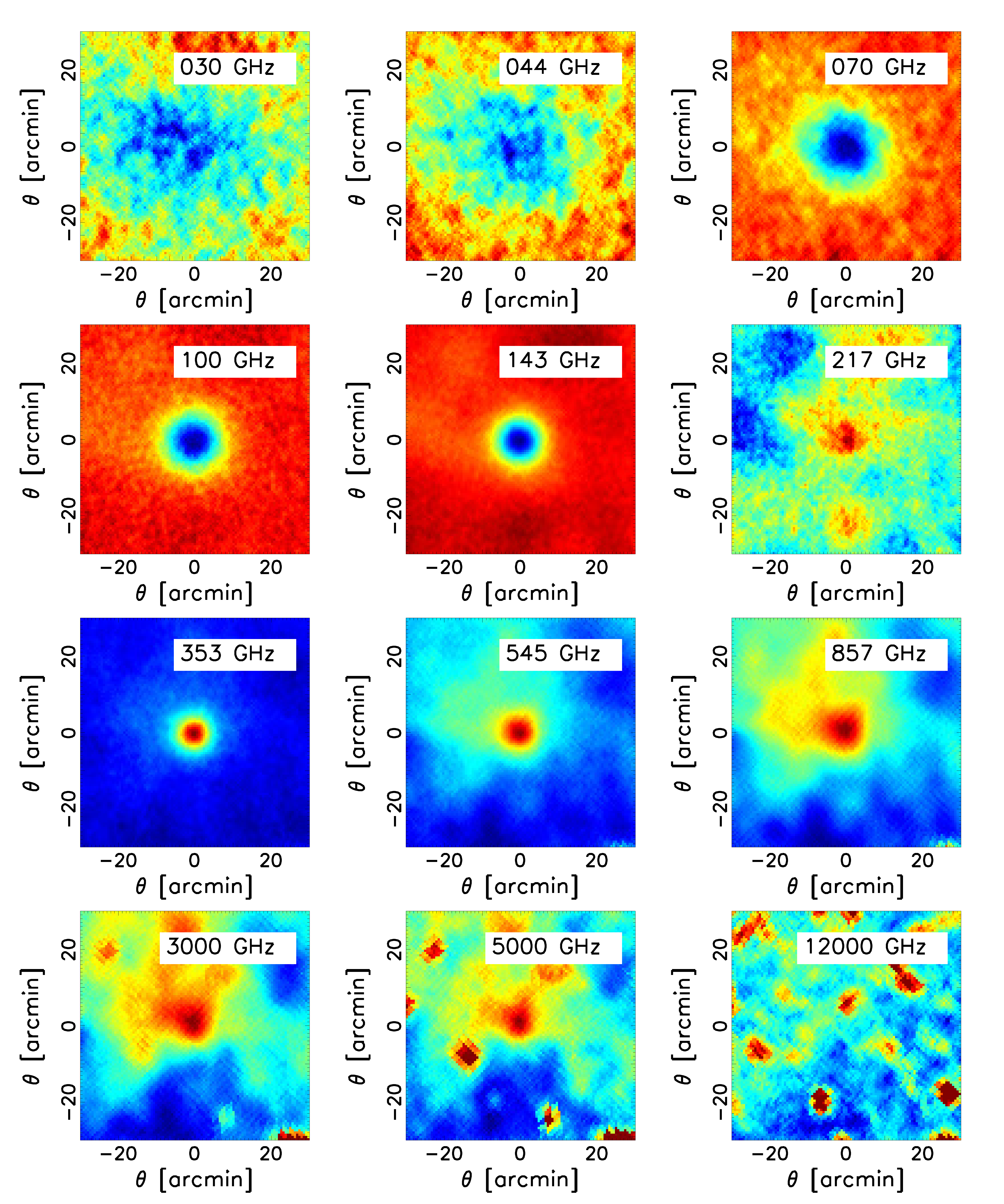}
\caption{From left to right and top to bottom: observed stacked intensity maps
at 30, 44, 70, 100, 143, 217, 353, 545, 857, 3000, 5000, and 12\,000\,GHz
at the positions of confirmed SZ clusters. All maps are at the native angular
resolution of the relevant \Planck\ channel. The measured intensity increases
from dark blue to red.}
\label{figstack}
\end{center}
\end{figure}

In Fig.~\ref{figstack}, we present the stacked signal at the positions
of the sample of confirmed SZ clusters in \Planck\ data from 30 to
857\,GHz and in IRIS maps from 100 to 25\,$\mu$m.  At low frequencies
(below 217\,GHz) we observe the typical tSZ intensity decrement.  However,
at 353 and 545\,GHz we see a mix of the positive tSZ signal and infrared
emission.  We also note that the infrared emission can also be observed at
217\,GHz where the tSZ effect is negligible.  We note the presence of 
significant infrared emission in the \Planck\ 857\,GHz channel, where the
tSZ signal is negligible. Similarly
we find a significant infrared signal in the IRAS 100 and 60\,$\mu$m bands.

\subsubsection{The SED of galaxy clusters}

\label{phot}
Each stacked map from 70\,GHz upwards is created at a resolution of
${\rm FWHM}=13\arcm$.\footnote{At 30\,GHz and 44\,GHz, we keep the native
resolution of the intensity maps, i.e., FWHM values of 32\farcm34 and
27\farcm12, respectively.}  We measure
the flux, $F(\nu)$, in the stacked patches from 30 to 853\,GHz through
aperture photometry within a radius of 20\arcm\ and compute the mean signal
in annuli ranging from 30\arcm\ to 60\arcm\ to estimate the surrounding
background level.  We use aperture photometry in order to obtain a
model-independent estimation of the total flux, without assuming a particular
shape for the galaxy cluster profile.  Thus, the flux is computed as 
\begin{equation}
\widehat{F}(\nu) = K_\nu \left(\sum_{r<20\arcm} A_{\nu,p}
 - \sum_{{30\arcm}< r<60\arcm} A_{\nu,p}
 \frac{N_{r<20\arcm}}{N_{{30\arcm}<r<60\arcm}}\right) \Delta \Omega,
\end{equation}
where $A_{\nu,p}$ is the pixel $p$ separated from the centre of the
map by a distance $r$ in the stacked map $A_{\nu}$, $\Delta \Omega$ is
the solid angle of one pixel, $N_{X}$ is the number of pixels that
follow the condition $X$, and $K_\nu$ is the bias for the aperture
photometry (equal to one except at 30 and 44\,GHz).  The 30 and 44\,GHz
channels have a lower angular resolution and thus the aperture photometry
does not measure all the signal.  We compute the factor $K_\nu$
by assuming that the physical signal at 143\,GHz has the same spatial
distribution as that at 30 or 44\,GHz:
\begin{equation}
K_{\nu} = \frac{\widehat{F}(143)}{\widehat{F}^*(143,\nu)},
\end{equation}
where the flux $\widehat{F}^*(143,\nu)$, is computed on the
143\, GHz stacked map after being set to the resolution of the 30 and 44\,GHz
channels.

\begin{figure}[!ht]
\begin{center}
\includegraphics[scale=0.20]{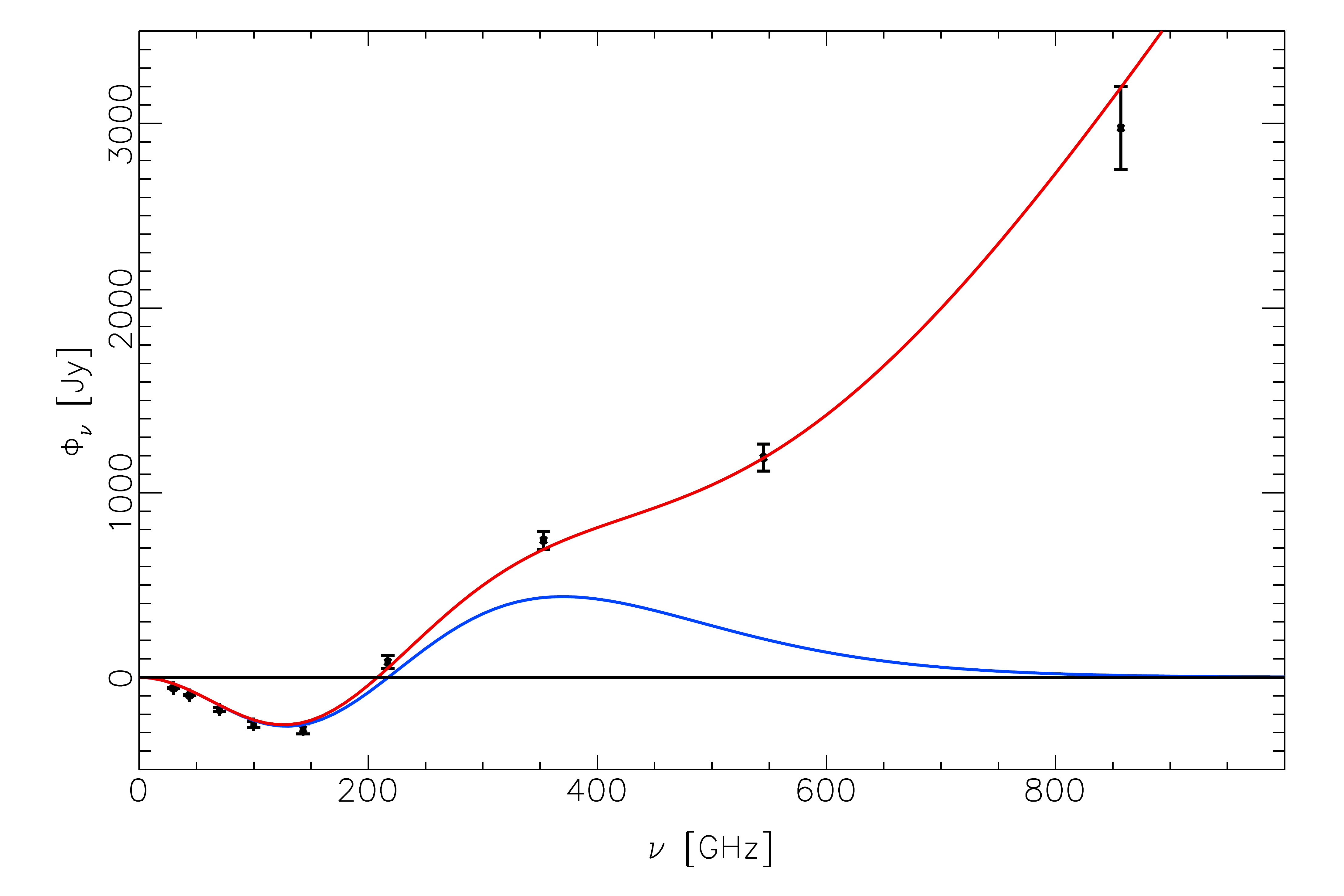}
\caption{Observed SED of the stacked signal towards galaxy clusters, from 30 to
  857\,GHz. In blue we show the tSZ contribution to the total SED
  and in red the total SED considering infrared emission.}
\label{figsed}
\end{center}
\end{figure}

From the scaling laws presented in Sect.~\ref{secmod1}, it is possible
to predict the expected SED of a cluster.  The total tSZ plus infrared emission
measured through stacking of $N_{\rm cl}$ clusters can be written as
\begin{equation}
F(\nu) = \sum_i^{N_{\rm cl}} \left[g(\nu) Y^{i}_{500}
 + S^{i}_{500}(\nu)\right],
\end{equation}
where ${S}_{500}(\nu) = a\, L_{500}(\nu)/4 \pi \chi^2(z)$ is
based on values from Table~\ref{tabscal} and we fit for $L_0$
(see Eq.~\ref{ciblm} for the $L_{500}$ expression), which
sets the global amplitude of the infrared emission in clusters.

\begin{figure}[!ht]
\begin{center}
\includegraphics[scale=0.20]{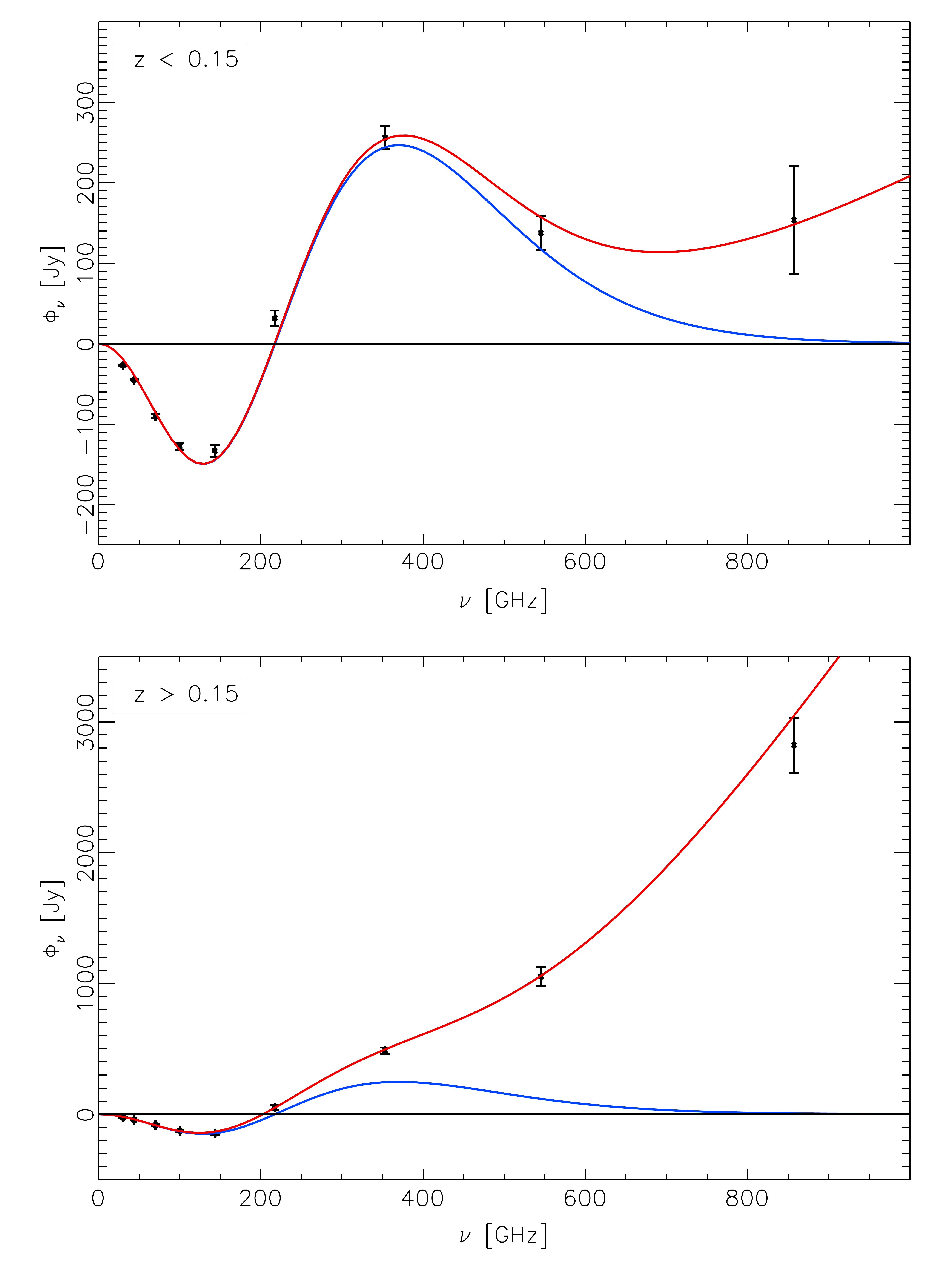}
\caption{Observed SED of the stacked signal toward galaxy clusters, from 30 to
  857\,GHz. {\it Top}: objects below $z=0.15$. {\it Bottom}:
  objects above $z=0.15$. In blue we show the tSZ contribution to
  the total SED and in red the total SED considering infrared
  emission.}
\label{figsedb}
\end{center}
\end{figure}

Figure~\ref{figsed} presents the derived SED towards galaxy clusters
compared to the tSZ-only SED. The observed flux at high frequencies, from
353 to 857\,GHz, calls for an extra infrared component to account for the
observed emission. We also present in Fig.~\ref{figsedb} the same SED for
two wide redshift bins,
below and above $z=0.15$, with median redshift $0.12$ and $0.34$,
respectively. We observe that most of the infrared
emission is produced by objects at $z > 0.15$.  We compare this
stacking analysis to the SED prediction from the scaling relation used to
reproduce results presented in Fig.~\ref{figsed} (red lines). This shows that
the modelling reproduces the observed redshift dependence of the infrared
flux of galaxy clusters.

\begin{figure}[!ht]
\begin{center}
\includegraphics[scale=0.17]{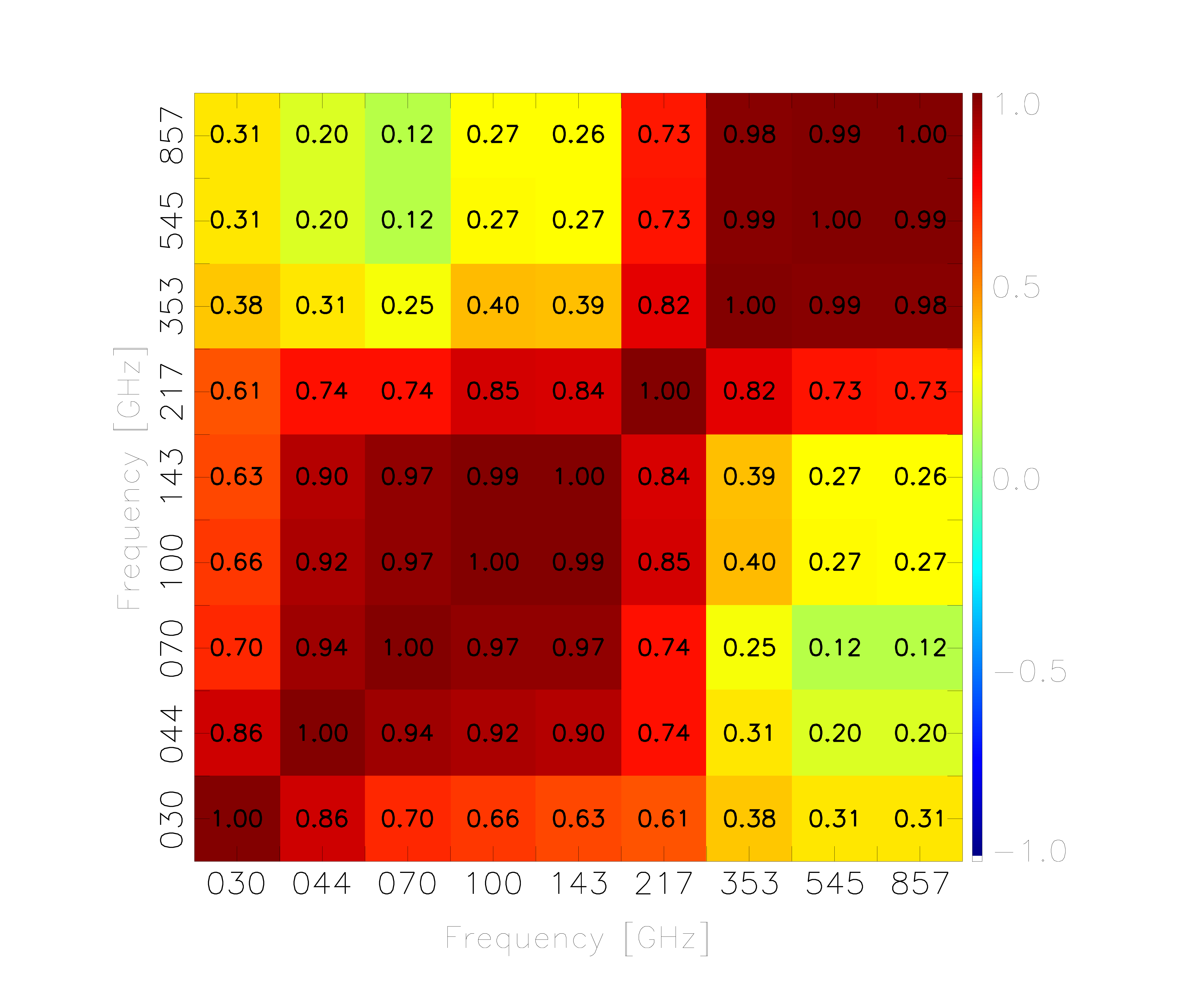}
\caption{Correlation matrix of ${\widehat F}(\nu)$ estimated from
  1000 random positions across the sky.}
\label{figcor}
\end{center}
\end{figure}

Uncertainties on $\widehat{F}(\nu)$ are mainly produced by
contamination from other astrophysical components. To estimate the
induced contamination level, we extract fluxes, $F'_{\nu,q}$, at 1000
random positions, $q$, across the sky with the same aperture
photometry as the one presented in Sect.~\ref{phot}. To derive a
realistic estimation of the noise, we avoid the Galactic plane area
for the random
positions, since this area is not represented in our cluster sample
(for details of the sky coverage of the PSZ1 and PSZ2
catalogues see \citealt{planck2013-p05a} and \citealt{planck2014-a36}).
In the stacking process each cluster is considered uncorrelated with the 
others. Indeed, considering the small number of objects, we can safely neglect
correlations induced by clustering in the galaxy cluster distribution.
The $F_{\nu}$ uncertainty correlation matrix is presented in
Fig.~\ref{figcor}. It only accounts for uncertainties produced by
uncorrelated components (with respect to the tSZ effect) in the flux
estimation.  From 44 to 217\,GHz, the CMB anisotropies are the main
source of uncertainties. This explains the high level of correlation
in the estimated fluxes. The 30\,GHz channel has a lower level of
correlation due to its high noise level.  At higher frequencies, from
353 to 857\,GHz, dust residuals becomes the dominant source in the
total uncertainty, which explains the low level of correlation with
low frequency channels. Contamination by radio sources inside galaxy clusters
can be neglected (at least statistically), since the measured fluxes at
30 and 44\,GHz agree with the tSZ SED.

\subsubsection{Comparison between tSZ, IR, and galaxy-number radial profiles}
\label{seccompa}

We computed the radial profile of each stack map from 70 to 857
GHz in native angular resolution.These profiles were calculated on a regular radial grid of annuli
with bins of width $\Delta r = 1\arcm$, allowing us to sample the stacked
map at a resolution similar to the \Planck\ pixel size.
The profile value in a bin is defined as the mean of the values of all
pixels falling in each annulus. We subtract a background offset
from the maps prior to the profile computation. The offset value is
estimated from the surrounding region of each cluster ($30\arcm < r <
60\arcm$).  The uncertainty associated with this baseline offset
subtraction is propagated into the uncertainty of each bin of the
radial profile.

\begin{figure}[!ht]
\begin{center}
\includegraphics[scale=0.20]{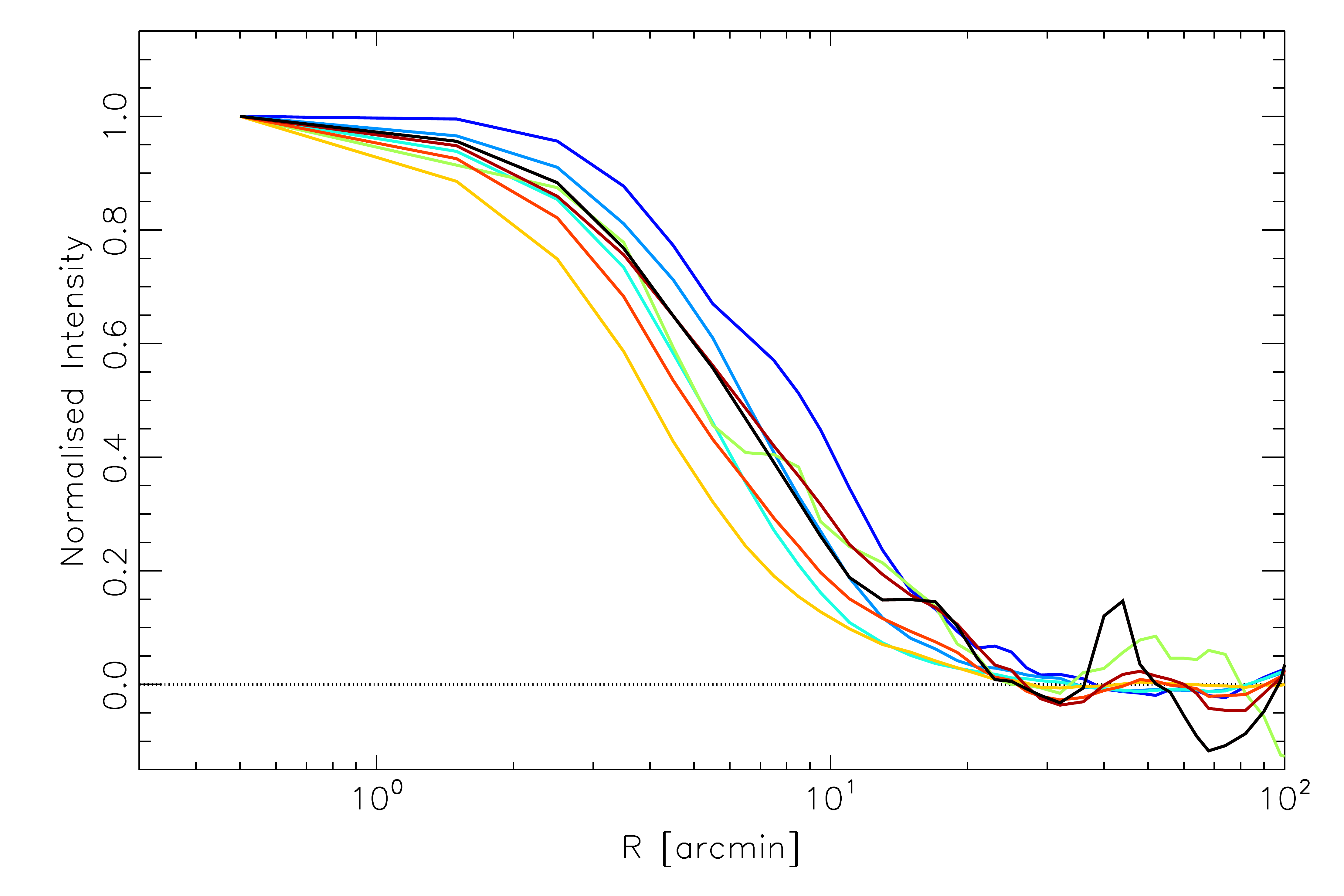}
\includegraphics[scale=0.20]{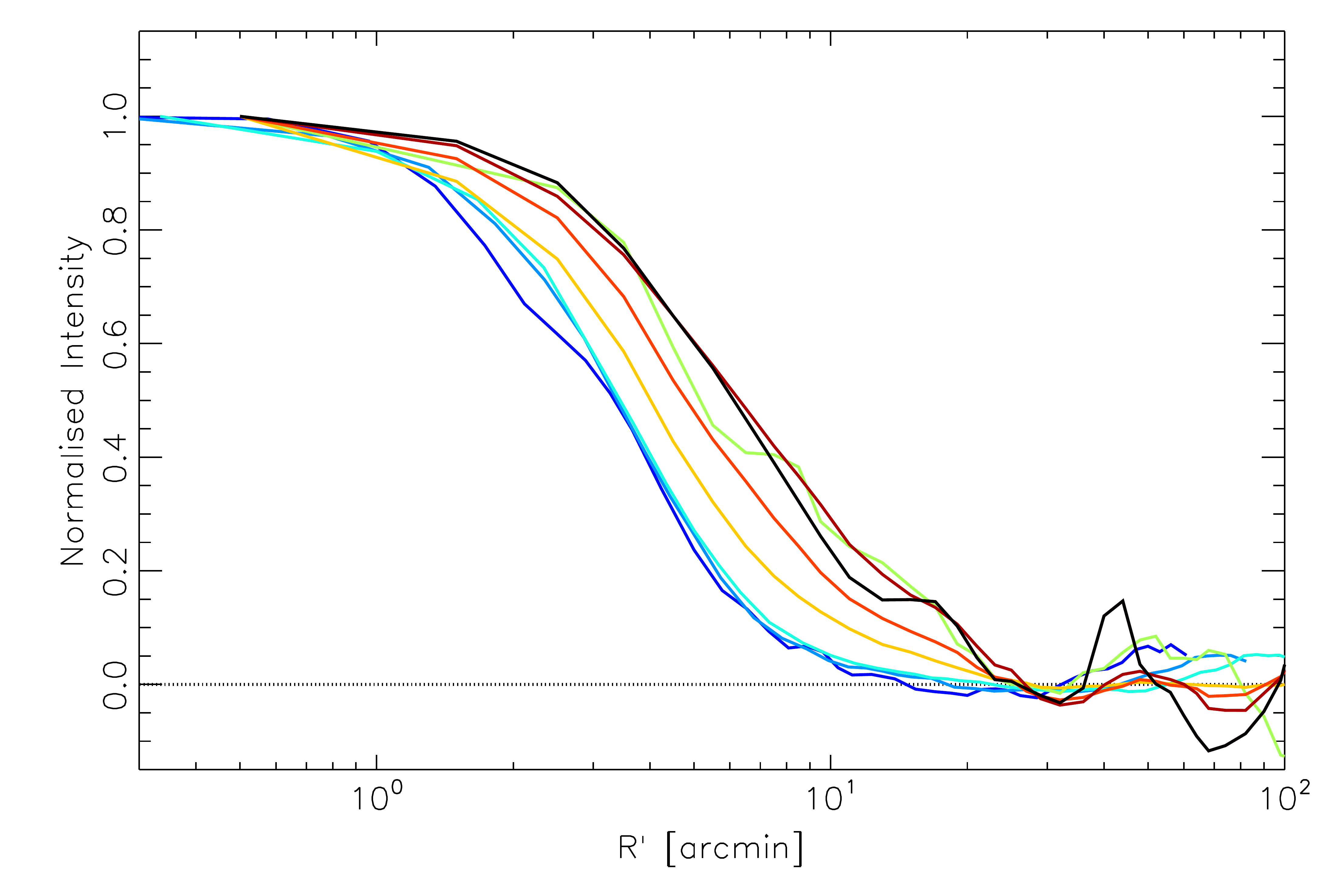}
\caption{{\it Top}: observed radial profile of the stacked signal toward galaxy
clusters at the native angular resolution at 70\,GHz (in dark blue), 100\,GHz
(blue), 143\,GHz (light blue), 217\,GHz (green), 353\,GHz (orange),
545\,GHz (red), 857\,GHz (dark red), and 100\,$\mu$m (black).
{\it Bottom}: same as top panel, but as a function of the rescaled radius
$R\arcm = R ({\cal B}_{\nu}/{\cal B}_{857})$.}
\label{figprof}
\end{center}
\end{figure}

In Fig.~\ref{figprof}, we present the profile normalized to one at the
centre. We observe that profiles derived from low-frequency maps
show a larger extension due to beam dilution. The smallest extension
of the signal is obtained for 353\,GHz, then it increases with
frequency.  For comparison we also display the profile at 100\,$\mu$m,
which shows the same extension as the profiles at 217 and 857\,GHz
where there is no significant tSZ emission.
For illustration, in the bottom panel of Fig.~\ref{figprof}, we display the
profiles as a function of the rescaled radius
$R\arcm = R ({\cal B}_{\nu}/{\cal B}_{857})$, with ${\cal B}_\nu$
the FWHM of the beam at frequency $\nu$.  Under the assumption that the tSZ
profile is Gaussian, this figure
allows us to directly compare the extension of the signal at all frequencies;
it illustrates the increase of the signal extension with frequency, except for
the 217\,GHz profiles, which have the same size as the high-frequency profiles.

We then define the extension of the \Planck\ profiles as ${\cal E}(\nu)$, where
\begin{equation}
{\cal E}^2(\nu) = 4\pi \, \ln2 \left(\frac{\int r p(r,\nu) {\rm d}r}
 {\int p(r,\nu) {\rm d}r} \right)^2 - {\cal B}^2_{\nu},
\end{equation}
with $p(r,\nu)$ the profile of the stacked signal at frequency $\nu$.
Integration of the
profiles is performed up to $r=30\arcm$.
The quantity ${\cal E}(\nu)$ is equivalent to a
FWHM for a Gaussian profile; however, we notice that the profile of
the stacked signal deviates from a Gaussian at large radii. This
increases the values obtained for the profile extension.  We estimate
the uncertainty on the profile extension using 1000 random positions
on the sky.

\begin{figure}[!ht]
\begin{center}
\includegraphics[scale=0.20]{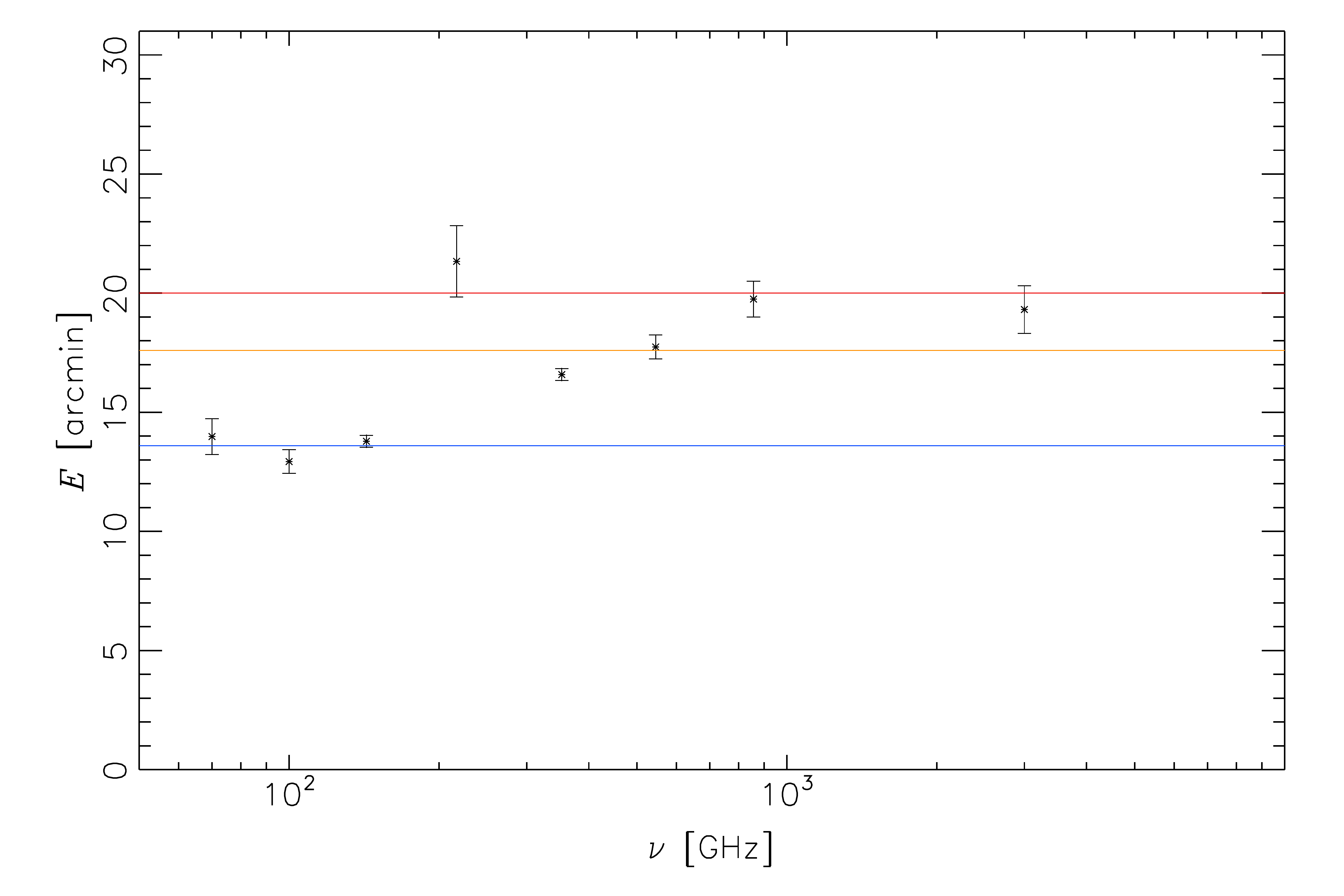}
\caption{Variation of the spatial extension, ${\cal E}(\nu)$, of the stacked
  signal from 70 to 3000\,GHz (100$\,\mu$m). The blue line shows the expected
  value for the GNFW profile from \citet{arn10} and $\theta_{500}$
  values from PSZ1, the orange line shows the value we derive with the
  \citet{xia11} profile, and the red line shows the value for an NFW
  profile with $c_{500} = 1.0$.}
\label{stackstd}
\end{center}
\end{figure}

In Fig.~\ref{stackstd}, we present the variation of ${\cal E}(\nu)$ as a
function of frequency from 70\,GHz  to 3000\,GHz (100$\,\mu$m).  We see lower
values of ${\cal E}(\nu)$ at low frequencies. The observed signal is composed
of two separate components, the tSZ effect and the infrared emission from
clusters.  The variation of ${\cal E}(\nu)$ is produced by the difference
between the spatial extension of the tSZ effect and the infrared emission. At
frequencies dominated by the tSZ signal (from 70 to 143\,GHz), we observe
${\cal E}(\nu) = 13\farcm6 \pm 0\farcm1$.  The expected value for the
\citet{arn10} GNFW profile and $\theta_{500}$ values for the \Planck\
clusters is ${\cal E}(\nu)=13\farcm6$. We observe
${\cal E}(\nu)=20\farcm0 \pm 0\farcm5$ at 217, 857, and
3000\,GHz. At 217\,GHz, where the tSZ signal is almost null, ${\cal E}(\nu)$
is dominated by the infrared emission and is similar to the signal found at
high frequencies (857\,GHz and 100\,$\mu$m). Considering an NFW profile
for infrared emission,
\begin{equation}
p(r) \propto \frac{1}{\left( c_{500}\frac{r}{R_{500}} \right)
 \left( 1 + c_{500}\frac{r}{R_{500}} \right)^2 }, 
\label{cibpro}
\end{equation}
and $\theta_{500}$ values for the \Planck\ clusters, the previous result
translates into constraints on $c_{500}$, giving
$c_{500} = 1.00^{+0.18}_{-0.15}$.  For comparison, we also display in
Fig.~\ref{stackstd} the prediction based on the profile used in
\citet{xia11}, which assumes a concentration 
\begin{equation}
c_{\rm vir} = \frac{9}{1+z}\left(\frac{M}{M_*}\right)^{-0.13},
\end{equation}
where $M_*$ is the mass for which $\nu(M,z) = \delta_{\rm c}/\left(D_{\rm g}
\sigma(M)\right)$ is equal to 1 \citep{bul01}, with
$\delta_{\rm c}$ the critical over-density,
$D_{\rm g}$ the linear growth factor, and $\sigma(M)$ the present-day rms
mass fluctuation.

This model leads to ${\cal E}(\nu)=17\farcm6$.  As a
consequence, our results demonstrate that galaxies in the outskirts of clusters
give a larger contribution to the total infrared flux than galaxies in the
cluster cores.  Indeed, star formation in outlying galaxies is not yet
completely quenched, while galaxies in the core no longer have a significant
star formation rate.  This result is consistent with previous
analyses that found radial dependence for star-forming
galaxies \citep[e.g.,][]{wei10,bra11,cop11,san13,muz14}.

We also model the  radial dependence of the average specific star-formation
rate (SSFR) as $\left[1 - A_{\rm q}{\rm exp} \left(-\alpha_{\rm q}
 r/R_{500}\right) \right]$, where $1-A_{\rm q}$ is the ratio
between the SSFR of a core galaxy and an outlying galaxy and
$\alpha_{\rm q}$ is the radial dependence of the infrared emission suppression.
We adopt the profile from \citet{xia11} for the galaxy distribution and,
using $A_{\rm q} \simeq 0.7$, we derive $\alpha_{\rm q} = 0.5^{+0.5}_{-0.2}$.

\subsection{Cross-correlation between the tSZ catalogue and temperature maps}
\label{seccat}

\begin{figure*}[!th]
\begin{center}
\includegraphics[scale=0.4]{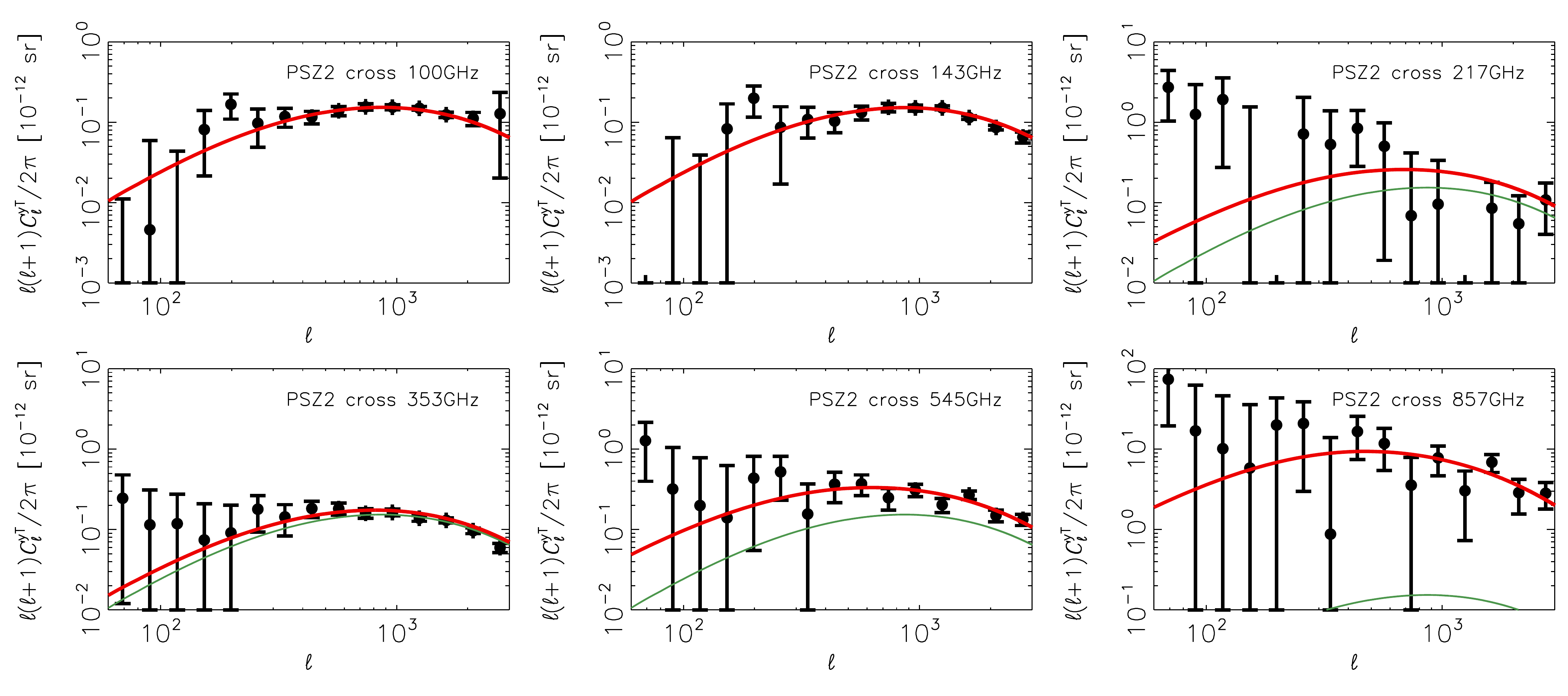}
\caption{Observed cross-correlation between the tSZ map ($y_{\rm c}$) of known
  clusters and the \Planck\ frequency maps ($T_\nu$), from 100 to 857\,GHz. 
  The data are presented as black circles, the tSZ
  auto-correlation is in green and the total model
  $y_{\rm c}$--$T_\nu$ accounting for tSZ-CIB correlation is in red
  for the best-fit of the CIB and tSZ spectra. All power spectra are
  presented in Compton-parameter units. Uncertainties are dominated by
  foreground residuals; thus they are highly correlated from one channel to
  another.}
\label{yf99mod}
\end{center}
\end{figure*}

\subsubsection{Methodology}

We focus on the detection of the correlation between
tSZ and infrared emission at the positions of confirmed galaxy clusters. To
do so, we use a map constructed from the projection of confirmed SZ
clusters on the sky, hereafter called the ``reprojected tSZ map,'' (see
end of Sect.~\ref{secdats}).  We measure the tSZ-CIB cross-correlation by
computing the angular cross-power spectrum, $C_\ell^{y_{\rm c} T_\nu}$,
between the reprojected tSZ map, $y_{\rm c}$, and the \Planck\ intensity maps,
$T_\nu$, from 100\,GHz to 857\,GHz. We note that $y_{\rm c}$ is only a fraction
of the total tSZ emission of the sky, $y$.  The cross-spectra are
\begin{equation}
C_\ell^{y_{\rm c} T_\nu} = g(\nu) C_\ell^{y_{\rm c},y}
 + C_\ell^{y_{\rm c}, {\rm CIB}(\nu)},
\label{modyc}
\end{equation}
where $C_\ell^{y_{\rm c}, {\rm CIB}(\nu)}$ is the cross-correlation between the
reprojected tSZ map (in Compton parameter units) and the CIB at frequency
$\nu$. Considering that the tSZ power spectrum is dominated by the 1-halo
term, we have $C_\ell^{y_{\rm c},y} \simeq C_\ell^{y_{\rm c},y_{\rm c}}$.
We mask the thermal dust emission from the Galaxy, keeping only the cleanest 40\,\% of
the sky. This mask is computed by thresholding the 857\,GHz
Planck full sky map at 30\arcm\ FWHM resolution. We verified that we derive
compatible results with 30 and 50\,\% of the sky.  We bin the cross-power
spectra and correct them for the beam and mask effects. In practice, the mixing
between multipoles induced by the mask is corrected by inverting the
mixing matrix $\tens{M}_{bb}$ between bins of multipoles
\citep[see][]{tri05}.

We estimate the uncertainties on the tSZ-CIB cross-spectra as
\begin{equation}
 \left(\Delta C_\ell^{y_{\rm c} T_\nu}\right)^2 = \frac{1}{(2\ell + 1)f_{\rm
     sky}}\left[ \left( C_\ell^{y_{\rm c} T_\nu}\right)^2 + C_\ell^{y_{\rm c}
     y_{\rm c}}C_\ell^{T_\nu T_\nu} \right].
\end{equation}
We stress that due to cosmic variance the uncertainties on the
$C_\ell^{y_{\rm c} T_\nu}$ spectra are highly correlated from frequency to
frequency.  We see in Fig.~\ref{yf99mod} that cross-spectra at
different frequencies show similar features. The covariance matrix
between cross-spectra at frequencies $\nu$ and $\nu'$ can be
expressed as
\begin{equation}
{\rm cov} \left(C_\ell^{y_{\rm c} T_\nu}, C_\ell^{y_{\rm c} T_{\nu'}}  \right)
 = \frac{1}{(2\ell + 1)f_{\rm sky}}
 \left[ C_\ell^{y_{\rm c} T_\nu}C_\ell^{y_{\rm c} T_{\nu'}}
 + C_\ell^{y_{\rm c} y_{\rm c}}C_\ell^{T_\nu T_{\nu'}} \right].
\end{equation}
Then we propagate the uncertainties through the bins and the mixing
matrix $\tens{M}^{-1}_{bb}$. We verify using Monte Carlo simulations that we
derive compatible levels of uncertainty.  In the following analysis, we
use the full covariance matrix between multipole bins.

We also consider uncertainties produced by the \Planck\ bandpasses
\citep{planck2013-p03d}. This source of uncertainty reaches up to
20\,\% for the tSZ transmission at 217\,GHz.  We also account for relative
calibration uncertainties \citep{planck2014-a09} ranging from
0.1\,\% to 5\,\% for different frequencies.
We verify that our methodology is not biased by systematic effects by
cross-correlating \Planck\ intensity maps with nominal cluster centres
randomly placed on the sky, and we observe a
cross-correlation signal compatible with zero.

\subsubsection{Results}
\label{secmes}

In Fig.~\ref{yf99mod}, we present the measured angular cross-power
spectra and our fiducial model. For convenience, all spectra are displayed
in Compton-parameter units; as a consequence the tSZ auto-correlation has the
same amplitude at all frequencies. We note that at 217\,GHz the tSZ
transmission is very faint. Thus, it induces large uncertainties when
displayed in Compton-parameter units; at this specific frequency, the
cross-power spectrum is dominated by the tSZ-CIB contribution, since the CIB
emission dominates over the tSZ emission at 217\,GHz ($g(\nu)$ becomes
negligible in Eq.~\ref{modyc}). Uncertainties are highly correlated from one
channel to another, explaining the similar features we observed in the
noise at all frequencies.

In order to address the significance of the tSZ-CIB correlation
in the measured cross-spectra we consider three cases to describe
the angular cross-power spectra and we compute the $\chi^2$ for
each case and each frequency:
\begin{itemize}
\item in Case~1, no tSZ auto-correlation and no tSZ-CIB correlation;
\item in Case~2, only a tSZ auto-correlation contribution;
\item in Case~3, both tSZ and tSZ-CIB spectra.
\end{itemize}
We present the derived $\chi^2$ values in Table~\ref{tabres}.

At low frequencies, 100 and 143\,GHz, the measured signal is completely
dominated by the tSZ auto-spectrum. The consistency between the
observed spectrum and the predicted one demonstrates that fluxes from
the \Planck\ SZ clusters are consistent with our measurement. At
intermediate multipoles the tSZ-CIB contribution has a similar
amplitude to the contribution of tSZ auto-spectra.  
At 217 and 353\,GHz we observe a higher value for the $\chi^2$ in
Case~3 compared to the value for in Case~2. However, the
difference between these values is not significant, considering the number
of degree of freedom per spectra. Thus the tSZ-CIB contribution is not
significant for these frequencies.  However,
at 545 and 857\,GHz we detect a significant excess with respect to the tSZ
auto-correlation contribution. We detect the tSZ-CIB contribution for 
low-redshift objects at 5.8 and 6.0$\,\sigma$ at 545 and 857\,GHz, respectively.

\begin{table}[!tbh]
\begingroup
\newdimen\tblskip \tblskip=5pt
\caption{Value of $\chi^2$ for the $y$--$T_\nu$ spectra from 100 to
857\,GHz (presented in Fig.~\ref{yf99mod}) coming from a null test
(Case~1), using only the tSZ spectra (Case~2), and when considering both
tSZ and tSZ--CIB spectra (Case~3).  The adjustment is performed in the
multipole range $200 > \ell > 2500$ with 60 degrees of freedom
(approximately 10 degrees of freedom per frequency). Thus each $\chi^2$ value
should be considered to be associated with $N_{\rm dof}\simeq10$.}
\label{tabres}
\nointerlineskip
\vskip -3mm
\footnotesize
\setbox\tablebox=\vbox{
 \newdimen\digitwidth
 \setbox0=\hbox{\rm 0}
  \digitwidth=\wd0
  \catcode`*=\active
  \def*{\kern\digitwidth}
  \newdimen\signwidth
  \setbox0=\hbox{+}
  \signwidth=\wd0
  \catcode`!=\active
  \def!{\kern\signwidth}
\halign{\hbox to 2.0cm{#\leaderfil}\tabskip=1.0em&
  \hfil#\hfil\tabskip=1.0em&
  \hfil#\hfil\tabskip=1.0em&
  \hfil#\hfil\tabskip=1.0em&
  \hfil#\hfil\tabskip=1.0em&
  \hfil#\hfil\tabskip=1.0em&
  \hfil#\hfil\tabskip=0pt\cr
\noalign{\doubleline}
\omit& \multispan6\hfil$\nu$ [GHz]\hfil\cr
\noalign{\vskip -3pt}
\omit& \multispan6\hrulefill\cr
\noalign{\vskip 3pt}
\omit$\chi^2$\hfil& 100& 143& 217& 353& 545& 857\cr
\noalign{\vskip 4pt\hrule\vskip 3pt}
Case 1& 600.0& 617.3& 7.9& 368.2& 172.0& 45.0\cr
Case 2& **4.5& **8.3& 6.8& **6.6& *41.1& 43.1\cr
Case 3& **4.5& **7.8& 9.1& **8.0& **6.9& *7.0\cr
\noalign{\vskip 4pt\hrule\vskip 3pt}}}
\endPlancktable
\endgroup
\end{table}


\section{The total tSZ-CIB cross-correlation}
\label{secres2}

In this section we investigate the all-sky tSZ-CIB
cross-correlation using two different approaches: (i) the tSZ-CIB cross-power
using a tSZ Compton-parameter map (Sect.~\ref{ymap}); and (ii) the tSZ-CIB
cross-power from a study of cross-spectra between \Planck\ frequencies
(Sect ~\ref{secfreq}).

\subsection{Constraints on the tSZ-CIB cross-correlation from tSZ
$y$-map/frequency maps cross-spectra }
\label{ymap}

This section presents the tSZ-CIB estimation using the cross-correlation
between a \Planck\ tSZ map and \Planck\ frequency maps.
Since the tSZ map contains CIB residuals, we carefully modelled 
these residuals in order to estimate the contribution from the tSZ-CIB
correlations.

\subsubsection{Methodology}

We compute the cross-power spectra between the \Planck\
frequency maps and a reconstructed $y$-map\footnote{Available from
\url{http://pla.esac.esa.int/pla/}\,.} derived from
component separation \citep[see][and references therein]{planck2014-a28}.
We choose the {\tt MILCA} map and check that there are no significant
differences with the {\tt NILC} map (both maps are described in \citealt{planck2014-a28}).
This cross-correlation can be decomposed into four terms:
\begin{equation}
C_\ell^{\widehat{y},T_\nu} = g(\nu)C_\ell^{y,y} + C_\ell^{y,{\rm CIB}(\nu)}
 + g(\nu)C_\ell^{y,y_{\rm CIB}} + C_\ell^{y_{\rm CIB},{\rm CIB}(\nu)},
\label{clycib}
\end{equation}
where $y_{\rm CIB}$ is the CIB contamination in the tSZ map. We
compute the uncertainties as
\begin{equation}
{\rm cov}(C_\ell^{\widehat{y},T_\nu},C_\ell^{\widehat{y},T_\nu'}) =
\frac{C_\ell^{\widehat{y},\widehat{y}} C_\ell^{T_\nu,T_\nu'} +
  C_\ell^{\widehat{y},T_\nu}C_\ell^{\widehat{y},T_\nu'}}{(2 \ell + 1)
  f_{\rm sky}},
\end{equation}
where $f_{\rm sky}$ is the fraction of the sky that is unmasked.
We bin the cross-power spectrum and deconvolve the beam and mask
effects, then we propagate uncertainties as described in
Sect.~\ref{seccat}.

The cross-correlations of a tSZ-map built from component-separation
algorithms and \Planck\ frequency maps are sensitive to both the tSZ
auto-correlation and tSZ-CIB cross-correlation. But this cross-correlation
also has a contribution produced by the CIB contamination to the tSZ map.
In particular, this contamination is, by construction, highly correlated with
the CIB signal in the frequency maps. 

\subsubsection{Estimation of CIB leakage in the tSZ map}

\begin{figure}[!th]
\begin{center}
\includegraphics[scale=0.2]{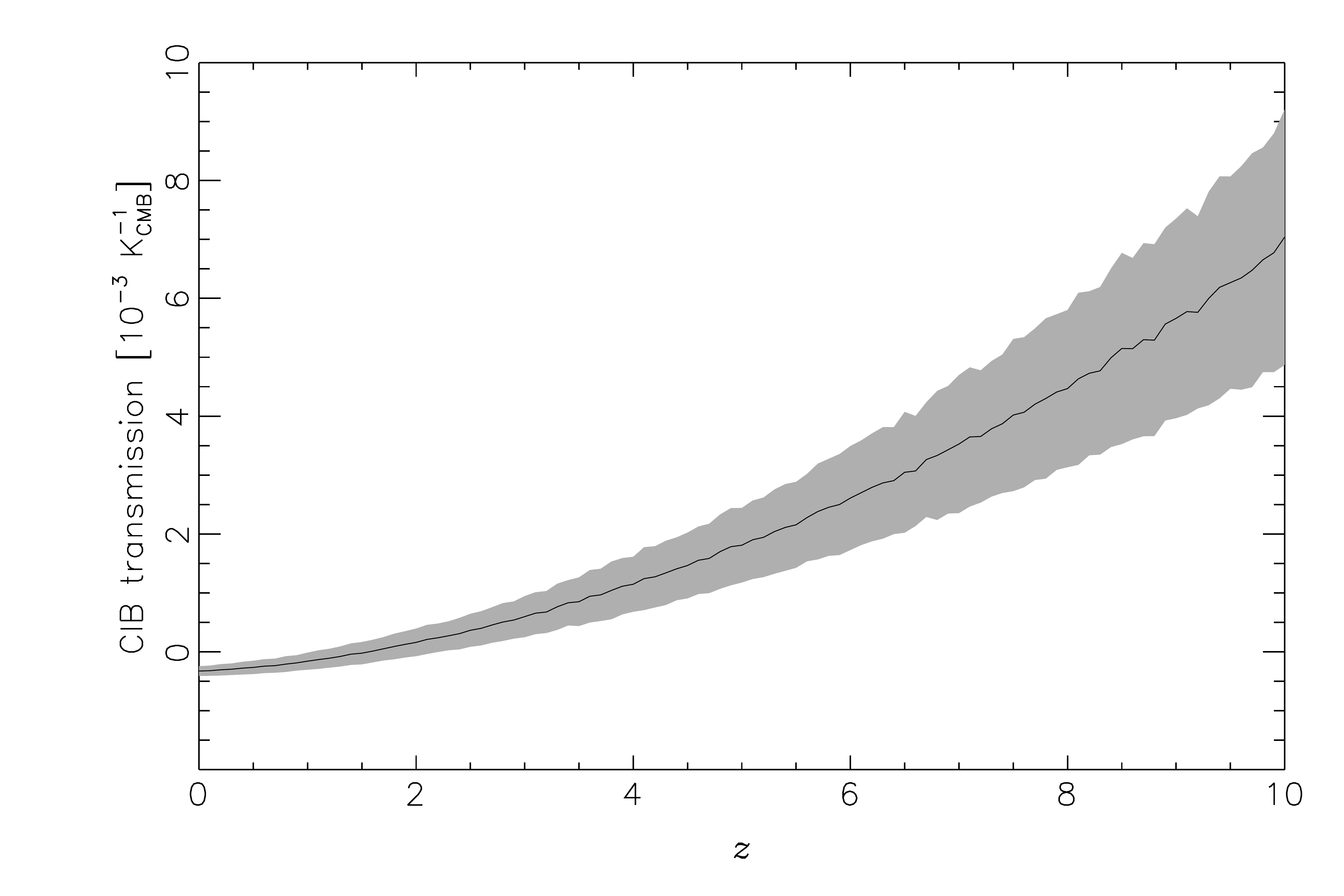}
\caption{Predicted CIB transmission in the tSZ $y$-map for a hypothetical CIB
emission of 1 K$_{\rm CMB}$ at 545\,GHz and at redshift $z$. The shaded area
represents the 68\% confidence region, for 
uncertainties of $\Delta T_{\rm d}=2\,$K and $\Delta \beta_{\rm d}=0.1$ in the
modified blackbody parameters of the hypothetical CIB emission.}
\label{cibleakz}
\end{center}
\end{figure}

The tSZ maps, denoted $\widehat{y}$ in the following, are derived using
component-separation methods. They are constructed through a linear
combination of \Planck\ frequency maps that depends on the angular scale
and the pixel, $p$, as
\begin{equation}
\widehat{y} = \sum_{i,j,\nu} w_{i,p,\nu} T_{i,p}(\nu).
\end{equation}
Here $T_{i,p}(\nu)$ is the \Planck\ map at frequency $\nu$ for the
angular filter $i$, and $w_{i,p,\nu}$ are the weights of the linear
combination.  Then, the CIB contamination in the $y$-map is
\begin{equation}
y_{\rm CIB} = \sum_{i,j,\nu} w_{i,p}(\nu) T^{\rm CIB}_{i,p}(\nu),
\end{equation}
where $T^{\rm CIB}(\nu)$ is the CIB emission at frequency $\nu$.
Using the weights $w_{i,p,\nu}$, and considering the CIB luminosity
function, it is possible to predict the expected CIB leakage as a
function of the redshift of the source by propagating the 
SED through the weights that are used to build the tSZ map.

\begin{figure}[!th]
\begin{center}
\includegraphics[scale=0.2]{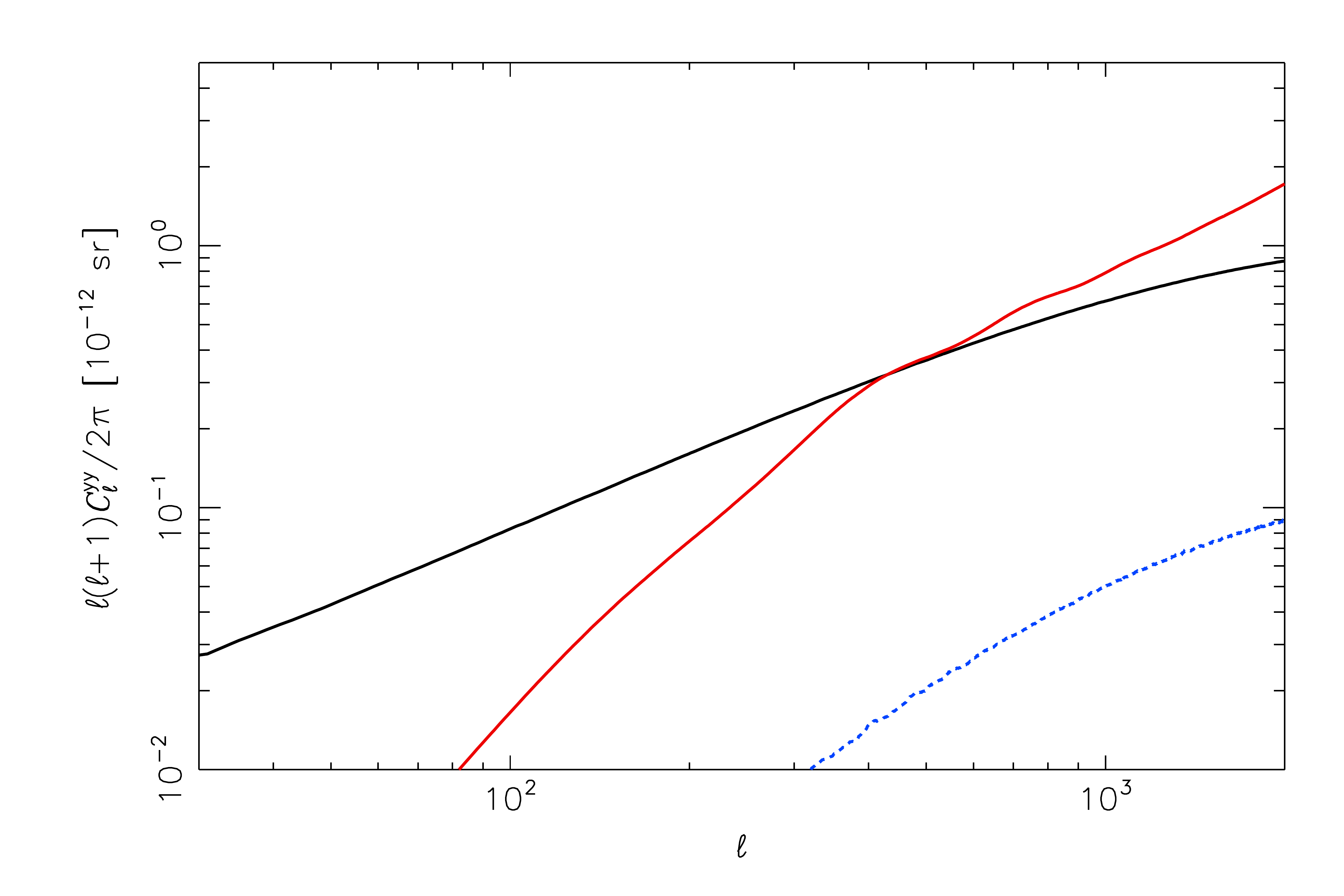}
\caption{Expected contribution to the tSZ power spectrum for the true tSZ
  signal (black curve), for CIB leakage (red curve), and for tSZ-CIB leakage
  contribution (blue curve). The dotted line indicates a negative
  power spectrum.}
\label{cibleakl}
\end{center}
\end{figure}

\begin{figure*}[!th]
\begin{center}
\includegraphics[scale=0.2]{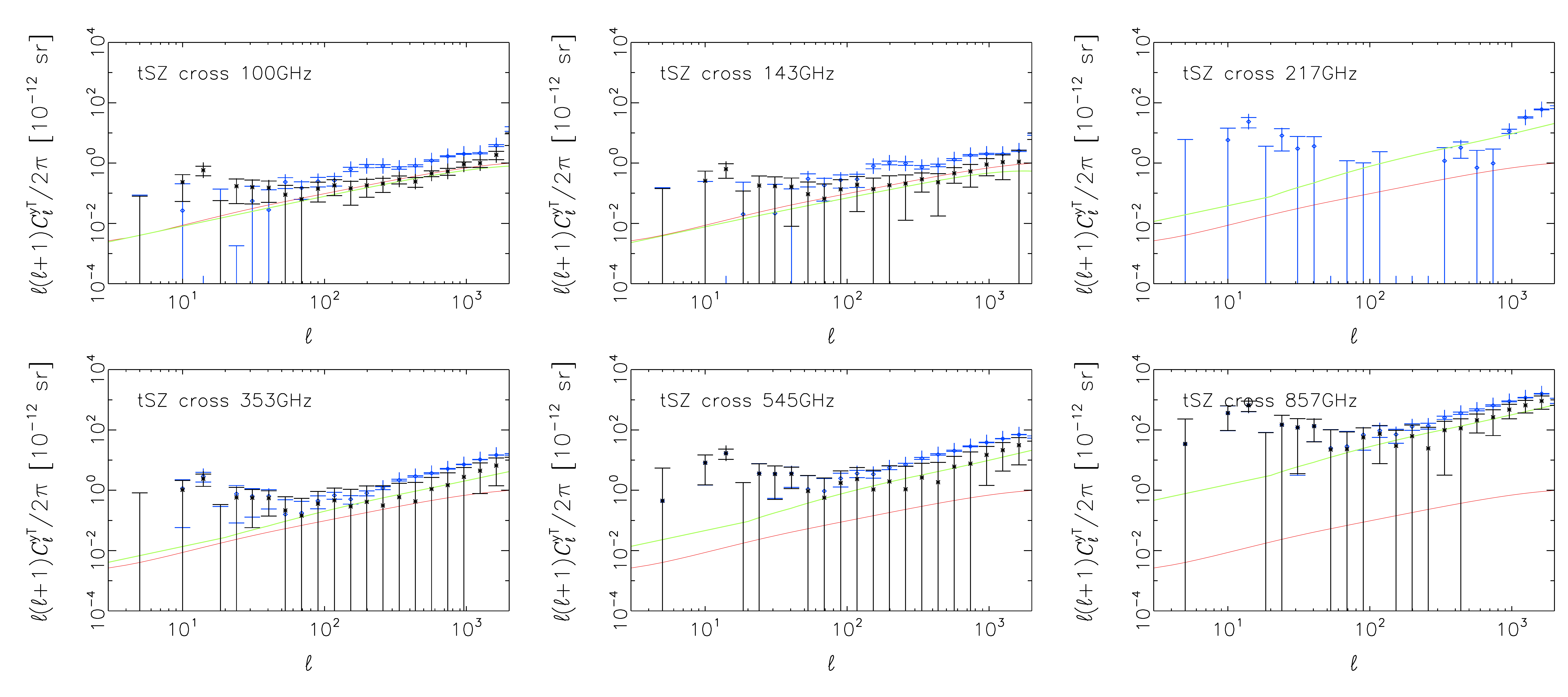}
\caption{From left to right and top to bottom: observed cross-correlation
  between the {\tt MILCA} $y$-map and the \Planck\ frequency maps from 30
  to 857\,GHz. In blue are the data points, in black the CIB-cleaned
  data points; the red solid line is the predicted signal from tSZ only
  and the green line is the total expected signal from the tSZ signal and
  the tSZ-CIB correlation. All spectra are presented in Compton-parameter
  units.}
\label{cibcibspeccont}
\end{center}
\end{figure*}

In Fig.~\ref{cibleakz}, we present the expected transmission of CIB
emission in the \Planck\ tSZ map for a $1\,{\rm K}_{\rm CMB}$ CIB source
at 545\,GHz at redshift $z$, based on the fiducial model for the
scaling relation presented in Sect.~\ref{secmod1}. 
The intensity of CIB leakage in the tSZ map is given by the integration of
the product of CIB transmission (presented in Fig.~\ref{cibleakz}) and
the CIB scaling relation at 545\,GHz.  Error bars account for SED
variation between sources. In this case we assume an
uncertainty of $\Delta T_{\rm d}=2\,$K and $\Delta \beta_{\rm d}=0.1$ on the
modified blackbody parameters.  We observe that the CIB at low $z$ leaks into
the tSZ map with only a small amplitude, whereas higher-redshift CIB produces
a higher, dominant, level of leakage. Indeed, ILC-based component-separation
methods tend to focus on Galactic thermal dust removal, and thus are less
efficient at subtracting high-$z$ CIB sources that have a different SED.

The CIB power spectra have been constrained in previous \Planck\
analyses \citep[see, e.g.,][]{planck2011-6.6,planck2013-pip56}, as presented in
Sect.~\ref{secauto}. We can use this knowledge of the CIB power spectra to
predict the expected CIB leakage, $y_{\rm CIB}$, in the tSZ map, $y$.
We performed 200 Monte Carlo simulations of tSZ and CIB maps that follow the
tSZ, CIB, and tSZ-CIB power spectra. Then, we applied to these simulations
the weights used to build the tSZ map.  Finally, we estimated the CIB
leakage and its correlation with the tSZ effect.  The tSZ map signal,
$\widehat{y}$, can be written as
$\widehat{y} = y + y_{\rm CIB}$. Thus, the spectrum of the tSZ map is
$C_\ell^{\widehat{y},\widehat{y}} = C_\ell^{y,y}
 + C_\ell^{y_{\rm CIB},y_{\rm CIB}} + 2 C_\ell^{y,y_{\rm CIB}}$.

In Fig.~\ref{cibleakl}, we present the predicted contributions to the tSZ
map's power spectrum for tSZ, CIB leakage, and tSZ-CIB leakage.  We
observe that at low $\ell$ (below 400) the tSZ signal dominates
CIB leakage and tSZ-CIB leakage contamination, whereas for higher $\ell$ the
signal is dominated by the CIB leakage part. The tSZ-CIB leakage spectrum
(dotted line in the figure) is negative, since it is dominated by low-$z$
($z\la2$) CIB leakage.

We also estimate the uncertainties on
$C_\ell^{y_{\rm CIB},y_{\rm CIB}}$, using the uncertainty on the CIB
correlation matrix from \citet{planck2013-pip56}.  We derive an average
uncertainty of 50\,\% on the CIB leakage amplitude in the tSZ map. This
uncertainty is correlated between multipoles at a level above 90\,\%.
Consequently, the uncertainty on the CIB leakage in the tSZ-map can be
modelled as an overall amplitude factor.

\subsubsection{Results}

By cross-correlating the simulated CIB leakage signal with the simulated CIB
at each frequency, it is also possible to predict the CIB leakage in the
cross-spectra between tSZ map and \Planck\ frequency maps. We can correct $C_\ell^{\widehat{y},T_\nu}$ spectra (Eq.~\ref{clycib})
using the estimated tSZ-CIB leakage cross-correlation term,
$g(\nu)C_\ell^{y,y_{\rm CIB}}$, and the CIB-CIB leakage
cross-correlation term, $C_\ell^{y_{\rm CIB},{\rm CIB}(\nu)}$, giving
\begin{equation}
  C_\ell^{\widehat{y},T_\nu,{\rm corr}}
  = C_\ell^{\widehat{y},T_\nu} - g(\nu)C_\ell^{y,y_{\rm CIB}}
  - C_\ell^{y_{\rm CIB},{\rm CIB}(\nu)}.
\end{equation}
Thus, the only remaining contributions are from the tSZ auto-correlation and
the tSZ-CIB cross-correlation.  We also propagate the associated
uncertainties.

Fig.~\ref{cibcibspeccont} shows the cross-correlation of the
tSZ-map and \Planck\ frequency maps after correcting for the
effects of the beam and mask, and for the terms
$g(\nu)C_\ell^{y,y_{\rm CIB}}$ and $C_\ell^{y_{\rm CIB},{\rm CIB}(\nu)}$.  As
was the case for Fig.~\ref{yf99mod}, all spectra are displayed in tSZ Compton-parameter
units, and the uncertainties present a high degree of correlation
from one frequency to another.  For each
cross-spectrum we adjust the amplitude, $A_{\rm tSZ-CIB}$, of the
tSZ-CIB contribution through a linear fit. The results of the fit are
listed in Table~\ref{tabcross}.  We obtain a maximum significance
of  2.3$\,\sigma$ at 857\,GHz and the results are consistent with the
fiducial model.

\begin{table}[!tbh]
\begingroup
\newdimen\tblskip \tblskip=5pt
\caption{Best-fit values for the tSZ--CIB amplitude, $A_{\rm tSZ-CIB}$,
using the fiducial model as reference.}
\label{tabcross}
\nointerlineskip
\vskip -3mm
\footnotesize
\setbox\tablebox=\vbox{
 \newdimen\digitwidth
 \setbox0=\hbox{\rm 0}
  \digitwidth=\wd0
  \catcode`*=\active
  \def*{\kern\digitwidth}
  \newdimen\signwidth
  \setbox0=\hbox{+}
  \signwidth=\wd0
  \catcode`!=\active
  \def!{\kern\signwidth}
\halign{\hbox to 2.0cm{#\leaderfil}\tabskip=1.5em&
  \hfil#\hfil\tabskip=1.5em&
  \hfil#\hfil\tabskip=0pt\cr
\noalign{\doubleline}
\omit$\nu$ [GHz]\hfil& $A_{\rm tSZ-CIB}$& $\Delta A_{\rm tSZ-CIB}$\cr
\noalign{\vskip 4pt\hrule\vskip 3pt}
100& $-3.6$& $3.8$\cr
143& $-1.6$& $3.7$\cr
353& $!2.0$& $2.0$\cr
545& $!1.7$& $1.4$\cr
857& $!1.6$& $0.7$\cr
\noalign{\vskip 4pt\hrule\vskip 3pt}}}
\endPlancktable
\endgroup
\end{table}

The combined constraints from 353, 545, and 857\,GHz measurements, as
well as the covariance structure of the measurement uncertainties, yield
an estimate of $A_{\rm tSZ-CIB} = 1.3 \pm 0.4$.  The uncertainties are
dominated by CIB leakage subtraction, which leads to highly correlated
uncertainties
of the 353, 545, and 857\,GHz spectra.  The optimal linear estimator we
use to calculate $A_{\rm tSZ-CIB}$ probes the measurements for the known
frequency trend of the CIB leakage in order to correct for this.  Since the
CIB leakage has affected all three of these correlated measurements in the
same direction, the combined
constraint of $1.3 \pm 0.4$ is below the range of the individual
(uncorrected) constraints ranging from 1.6 to 2.0.  Table~\ref{tabcross}
contains the necessary covariance information used here.

\subsection{Constraints on tSZ-CIB cross-correlation from \Planck\ frequency
maps}
\label{secfreq}
As a last approach, we explore the direct cross-correlation between \Planck\
frequency maps.

\subsubsection{Methodology}

In terms of tSZ and CIB components the cross-spectra between
frequencies $\nu$ and $\nu'$ can be written as
\begin{align}
    C_\ell^{\nu,\nu'} =& \,g(\nu)g(\nu')C^{\rm y,y}_\ell +
    C_\ell^{\rm CIB(\nu),CIB(\nu')}+ g(\nu)C_\ell^{\rm y,CIB(\nu')} \nonumber \\
    & + g(\nu')C_\ell^{\rm y,CIB(\nu)} + C_\ell^{\rm other}(\nu),
\end{align} 
where $C_\ell^{\rm other}(\nu)$ accounts for the contribution of all
components except for tSZ and CIB.
We compute the cross-spectra between \Planck\ frequency maps from
100 to 857\,GHz as
\begin{equation}
  C^{\nu,\nu'}_\ell = \frac{C^{\nu_1,\nu'_2}_\ell + C^{\nu_2,\nu'_1}_\ell}{2},
\end{equation}
where subscripts $1$ and $2$ label the ``half-ring'' \Planck\
maps. This process allows us to produce power spectra without the noise
contribution.  We also compute the covariance between spectra as
\begin{align}
{\rm cov}(C^{\nu,\nu'}_\ell, C^{\nu'',\nu'''}_\ell) &=
 \frac{C^{\nu_1,\nu''_1}_\ell C^{\nu'_2,\nu'''_2}_\ell
 + C^{\nu_1,\nu'''_2}_\ell C^{\nu'_2,\nu''_1}_\ell}{4(2\ell+1)f_{\rm sky}}
 \nonumber \\
&+ \frac{C^{\nu_1,\nu''_2}_\ell C^{\nu'_2,\nu'''_1}_\ell
 + C^{\nu_1,\nu'''_1}_\ell C^{\nu'_2,\nu''_2}_\ell}{4(2\ell+1)f_{\rm sky}}
 \nonumber \\
&+ \frac{C^{\nu_2,\nu''_1}_\ell C^{\nu'_1,\nu'''_2}_\ell
 + C^{\nu_2,\nu'''_2}_\ell C^{\nu'_1,\nu''_1}_\ell}{4(2\ell+1)f_{\rm sky}}
 \nonumber \\
&+ \frac{C^{\nu_2,\nu''_2}_\ell C^{\nu'_1,\nu'''_1}_\ell
 + C^{\nu_2,\nu'''_1}_\ell C^{\nu'_1,\nu''_2}_\ell}{4(2\ell+1)f_{\rm sky}}.
\end{align}
We correct the cross-spectra for beam and mask effects, using
the same Galactic mask as in Sect.~\ref{seccat}, removing 60\,\% of the sky,
and we propagate uncertainties on cross-power spectra
as described in Sect.~\ref{seccat}.

The tSZ and CIB contributions to $C_\ell^{\nu,\nu'}$ are contaminated by other
astrophysical emission.  We remove the CMB contribution in
$C_\ell^{\nu,\nu'}$ using the \Planck\ best-fit cosmology
\citep{planck2014-a15}.  We note that the \Planck\ CMB maps suffer from
tSZ and CIB residuals, so they cannot be used for our purpose.

\subsubsection{Estimation of tSZ-CIB amplitude}

We fit thermal dust, radio sources, tSZ, CIB (that accounts for the total
fluctuations in extragalactic infrared emission), and tSZ-CIB amplitudes,
$A_{\rm dust }$, $A_{\rm rad}$, $A_{\rm tSZ }$, $A_{\rm CIB }$,
and $A_{\rm tSZ-CIB }$, respectively, through a linear fit.
For the dust spectrum we assume $C_\ell \propto \ell^{\,-2.8}$
\citep{planck2013-pip56}, and for radio sources $C_\ell \propto \ell^{\,0}$.
For the tSZ-CIB correlation, tSZ, and CIB power spectra, we use templates
computed as presented in Sect.~\ref{secmod2}.  This gives us
\begin{align}
\label{eqmodmult}
C_\ell^{\nu,\nu'}  & = A_{\rm tSZ }g(\nu)g(\nu')C^{\rm y,y}_\ell \nonumber \\
	&+A_{\rm CIB} C_\ell^{\rm CIB(\nu),CIB(\nu')} \nonumber \\
	&+A_{\rm tSZ-CIB} \left[ g(\nu)C_\ell^{\rm y,CIB(\nu')}
 + g(\nu')C_\ell^{\rm y,CIB(\nu)}\right]\nonumber \\
	&+A_{\rm dust} f_{\rm dust}(\nu)f_{\rm dust}(\nu') \ell^{-2.8}
 \nonumber \\
	&+A_{\rm rad} f_{\rm rad}(\nu)f_{\rm rad}(\nu').
\end{align}
Here $f_{\rm dust}$ and $f_{\rm rad}$ give the frequency dependence of thermal
dust and radio point sources, respectively.
For thermal dust we assume a modified
blackbody emission law, with $\beta_{\rm d}=1.55$ and $T_{\rm d}=20.8\,$K
\citep{planck2013-p06b}.  For radio point sources we assume a spectral
index $\alpha_{\rm r}=-0.7$ \citep{planck2013-p05}.
The adjustment of the amplitudes and the estimation of the amplitude
covariance matrix are performed simultaneously on the six auto-spectra and
15 cross-spectra from $\ell=50$ to $\ell=2000$.

\begin{figure}[!th]
\begin{center}
\includegraphics[scale=0.18]{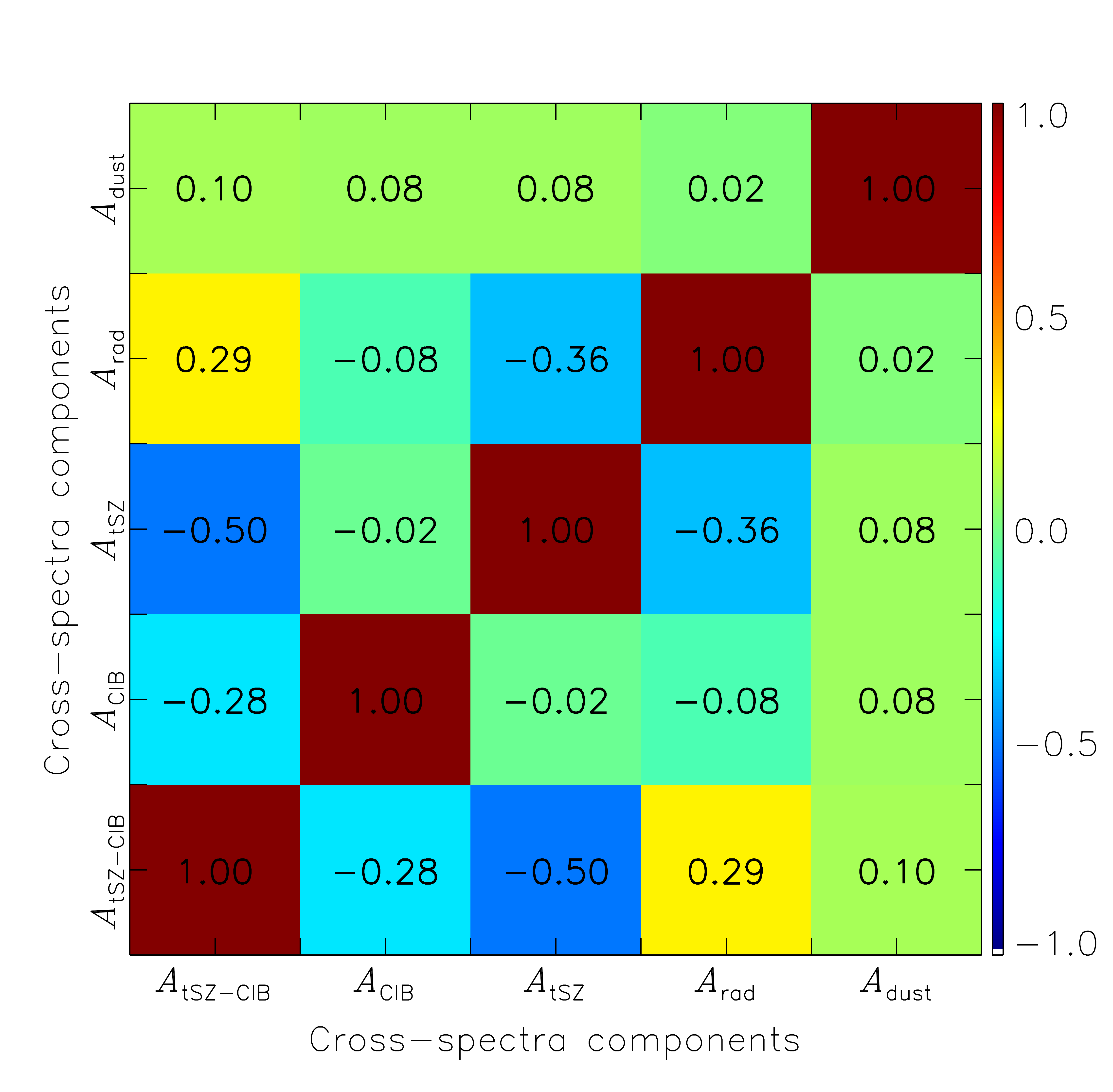}
\caption{Correlation matrix for multi-frequency cross-spectra components from
Eq.~(\ref{eqmodmult}).}
\label{plckcrosscormat}
\end{center}
\end{figure}

For tSZ and CIB spectra we reconstruct amplitudes compatible with
previous constraints (see Sect.~\ref{secauto}): $A_{\rm CIB} = 0.98 \pm 0.03$
for the CIB; and $A_{\rm tSZ} = 1.01 \pm 0.05$ for tSZ.  For the tSZ-CIB
contribution we obtain $A_{\rm tSZ-CIB} = 1.19 \pm 0.30$. Thus, we obtain a
detection of the tSZ-CIB cross-correlation at 4$\,\sigma$, consistent
with the model.  In Fig.~\ref{plckcrosscormat}, we present the
correlation matrix between cross-spectra component amplitudes. The
highest degeneracy occurs between tSZ and tSZ-CIB amplitudes, with a
correlation of $-50$\,\%. We also note that CIB and radio contributions
are significantly degenerate, with tSZ-CIB correlation amplitudes
of $-28$\,\% and 29\,\%, respectively.


\section{Conclusions and discussion}
\label{seccon}

We have performed a comprehensive analysis of the infrared emission
from galaxy clusters.  We have proposed a model of the
tSZ-CIB correlation derived from coherent modelling of both the tSZ and CIB at
galaxy clusters.  We have shown that the models of the tSZ and
CIB power spectra reproduce fairly well the observed power spectra from the
\Planck\ data.  Using this approach, we have been able to predict
the expected tSZ-CIB cross-spectrum.  Our predictions are consistent
with previous work reported in the literature \citep{add12,zah12}.

We have demonstrated that the CIB scaling relation
from \citet{planck2013-pip56} is able to reproduce the observed stacked
SED of \Planck\ confirmed clusters.  We have also set constraints on
the profile of the this emission and found that the infrared
emission is more extended than the tSZ profile. 
We also find that
the infrared profile is more extended than seen in previous work \cite[see
e.g.,][]{xia11} based on numerical simulation \citep{bul01}.
Fitting for the concentration of an NFW profile, the infrared emission shows
$c_{500} = 1.00^{+0.18}_{-0.15}$.  This demonstrates that the infrared
brightness of cluster-core galaxies is lower than that of outlying galaxies.

We have presented three distinct approaches for constraining the tSZ-CIB
cross-correlation level: (i) using confirmed tSZ clusters; (ii) through
cross-correlating a tSZ Compton parameter map with \Planck\
frequency maps; and (iii) by directly cross-correlating \Planck\
frequency maps.  We have compared these analyses with the predictions
from the model and derived consistent results.  We obtain:
(i) a detection of the tSZ-IR correlation at 6$\,\sigma$;
(ii) an amplitude $A_{\rm tSZ-CIB} = 1.5 \pm 0.5$; and (iii) an
amplitude $A_{\rm tSZ-CIB} = 1.2 \pm 0.3$.  At 143\,GHz these values
correspond to correlation coefficients at $\ell=3000$ of: (ii) $\rho =
0.18 \pm 0.07$; and (iii) $\rho = 0.16 \pm 0.04$.  These results are
consistent with previous analyses by the ACT collaboration, which set
upper limits $\rho < 0.2$ \citep{dun13} and by the SPT collaboration,
which found $\rho = 0.11^{+0.06}_{-0.05}$ \citep{geo14}.

Our results, with a detection of the full tSZ-CIB cross-correlation
amplitude at $4\,\sigma$, provide the tightest constraint so
far on the tSZ-CIB correlation factor.
Such constraints on the tSZ-CIB cross-correlation are needed to perform an
accurate measurement of the tSZ power spectrum.
Beyond power spectrum analyses, the tSZ-CIB cross-correlation is also a major
issue for relativistic tSZ studies, since CIB emission toward galaxy clusters
mimics the relativistic tSZ correction and thus could produce significant bias
if not accounted for properly.
This $4\,\sigma$ measurement of the amplitude of the tSZ-CIB correlation will
also be important for estimates of the
``kinetic'' SZ power spectrum.

\begin{acknowledgements}
The Planck Collaboration acknowledges the support of: ESA; CNES, and
CNRS/INSU-IN2P3-INP (France); ASI, CNR, and INAF (Italy); NASA and DoE
(USA); STFC and UKSA (UK); CSIC, MINECO, JA and RES (Spain); Tekes, AoF,
and CSC (Finland); DLR and MPG (Germany); CSA (Canada); DTU Space
(Denmark); SER/SSO (Switzerland); RCN (Norway); SFI (Ireland);
FCT/MCTES (Portugal); ERC and PRACE (EU). A description of the Planck
Collaboration and a list of its members, indicating which technical
or scientific activities they have been involved in, can be found at
\url{http://www.cosmos.esa.int/web/planck/planck-collaboration}.
Some of the results in this paper have been derived using the
{\tt HEALPix} package. We also acknowledge the support of
the French Agence Nationale de la Recherche under grant
ANR-11-BD56-015.
\end{acknowledgements}

\bibliographystyle{aat}
\bibliography{powspec_szcib_ter.bib,planck_bib.bib}

\end{document}